\documentclass[twocolumn]{IEEEtran}
\usepackage{epsfig,amsmath,amstext,amsfonts,amssymb,cite,cases,multirow}
\usepackage{algorithm} 
\usepackage{algorithmic} 
\usepackage{mathrsfs}
\usepackage{color}

\usepackage{amsmath}
\usepackage{graphicx}
\usepackage{subfigure}
\usepackage{bm}
\usepackage{subeqnarray}
\usepackage{cases}
\usepackage{diagbox}
\usepackage{makecell}

\begin{document}
	\title{PAPR Reduction Using Iterative Clipping/Filtering and ADMM Approaches for OFDM-Based Mixed-Numerology Systems}
	
	\author{
		Xiaoran~Liu,~
		Xiaoying~Zhang,~
		Lei~Zhang,~\IEEEmembership{Senior Member,~IEEE,}
		Pei~Xiao,~\IEEEmembership{Senior Member,~IEEE,}
		Jibo~Wei,~\IEEEmembership{Member,~IEEE,}
		Haijun~Zhang,~\IEEEmembership{Senior Member,~IEEE,}
		and Victor~C.~M.~Leung,~\IEEEmembership{Fellow,~IEEE}

			\thanks{X. Liu, X. Zhang and J. Wei are with the Department of Electronic Science, National University of Defense Technology, Changsha 410073, China (e-mail: \{liuxiaoran10, zhangxiaoying, wjbhw\}@nudt.edu.cn).				
			
			L. Zhang is with the School of Engineering, University of Glasgow, Glasgow, G12 8QQ, U.K.(e-mail: lei.zhang@glasgow.ac.uk).
			
			X. Pei is with the 5G Innovation Centre, and the Institute for Communication Systems, University of Surrey, Guildford GU2 7XH, U.K.(e-mail: p.xiao@surrey.ac.uk).
			
			H. Zhang is with the Beijing Advanced Innovation Center for Materials Genome Engineering, Beijing Engineering and Technology Research Center for Convergence Networks and Ubiquitous Services, University of Science and Technology Beijing, Beijing 100083, China (e-mail: haijunzhang@ieee.org).
			
			V. C. M. Leung is with the Department of Electrical and Computer Engineering, The University of British Columbia, Vancouver, BC V6T 1Z4, Canada (e-mail: vleung@ece.ubc.ca).

	}
		
	}
	
\maketitle
	
\begin{abstract}
Mixed-numerology transmission is proposed to support a variety of communication scenarios with diverse requirements. 
However, as the orthogonal frequency division multiplexing (OFDM) remains as the basic waveform, the peak-to average power ratio (PAPR) problem is still cumbersome. 
In this paper, based on the iterative clipping and filtering (ICF) and optimization methods, we investigate the PAPR reduction in the mixed-numerology systems. 
We first illustrate that the direct extension of classical ICF brings about the accumulation of inter-numerology interference (INI) due to the repeated execution. By exploiting the clipping noise rather than the clipped signal, the noise-shaped ICF (NS-ICF) method is then proposed without increasing the INI. 
Next, we address the in-band distortion minimization problem subject to the PAPR constraint. 
By reformulation, the resulting model is separable in both the objective function and the constraints, and well suited for the alternating direction method of multipliers (ADMM) approach. 
The ADMM-based algorithms are then developed to split the original problem into several subproblems which can be easily solved with closed-form solutions. 
{\color{black}
Furthermore, the applications of the proposed PAPR reduction methods combined with filtering and windowing techniques are also shown to be effective.
}
\end{abstract}
	
	
\section{Introduction}\label{sec1}
Future wireless communication systems will address unprecedented challenges to cope with a high degree of heterogeneity 
in terms of services, applications, {\color{black}use} cases, deployment scenarios and mobility levels \cite{2017ICM-ZhangMulti,2018ICSM-ZaidiOFDM}. 
In order to support multiform families of communication scenarios and applications, enhanced mobile broadband (eMBB), massive machine type communications (mMTC) and ultra reliable and low latency communications (URLLC) have been categorized as three major usage scenarios for the 5G cellular wireless communication.
However, these scenarios with diverse technical requirements generally lead to different physical layer {\color{black}designs}.
For example, 
the subcarrier spacing in mMTC scenario is expected to be small in order to support a massive number of devices with limited spectrum resource\cite{2016IA-A.IjazEnabling}. 
URLLC, on the other hand, designed for communication between devices/machines with high reliability and low end-to-end delay, can find its applications in vehicular communication, factory automation, remote surgery, etc\cite{2016-TR38.913Study}, 
and large subcarrier spacing (and accordingly, smaller symbol duration) is thus preferred for the stringent latency.
Therefore, the traditional ``one-fit-all" numerology design cannot simultaneously satisfy all these requirements\cite{2016IJSAC-C.L.INew}.

The Third  Generation Partnership Project (3GPP) has discussed the 5G new radio (NR) access technology that provides guidelines for the design of waveform and numerology\cite{2017-3GPPTechnical}.
Mixed-numerology is one of the important features proposed to efficiently support multiple services on the same carrier but with different subcarrier spacing of the orthogonal frequency division multiplexing (OFDM).
Compared with multiplexing in the time domain \cite{2012Sahin}, aligning different numerologies in frequency domain {\color{black}has} better forward compatibility and inclusive support for different latency services\cite{2017IToWC-ZhangSubband}.
Therefore, the frequency domain multiplexing that divides the system bandwidth into several subbands and assigns different numerologies to each subband has received widespread support\cite{2017IJoSAiC-Guan5G}.

However, multiplexing different OFDM numerologies in frequency domain gives rise to several new challenges in {\color{black}system design,
such as inter-numerology interference\cite{2018JoMM-YazarFlexible} and numerology scheduling\cite{2019EJoWCaN-YazarReliability}.
Additionally, high peak-to-average-power ratio (PAPR) inherited from multicarrier modulation is still a major drawback in OFDM-based mixed-numerology systems.}
In order to avoid the nonlinear distortion caused by imperfect power amplifier (PA)\cite{2009JCaN-D.-W.LimOverview},
the PAPR problem should be carefully addressed for mixed-numerology systems\cite{2017ICM-Lien5G}.

\subsection{Related Works}
For the conventional OFDM system, a variety of PAPR reduction techniques have been proposed over the past decades.
Those techniques can be roughly classified into three categories, i.e., multiple signaling probabilistic,
 coding techniques, and signal distortion\cite{2013ICST-RahmatallahPeak}. 
The first category is able to generate several candidate permutations of the OFDM signal and choose the one with the lowest PAPR, whose representative techniques include selective mapping (SLM)\cite{1996EL-BaumlReducing}, partial transmit sequence (PTS)\cite{1997EL-MullerOFDM}, and tone reservation (TR)\cite{1998Tellado}.
The coding techniques modifies its inherence to provide both the capability of error detection/correction and PAPR reduction\cite{2011ITC-M.SabbaghianShannon}.

Compared with other PAPR reduction techniques, the distortion based techniques can be implemented transparently to the existing standards, which is more attractive in current communication systems\cite{2016ICM-ZhangWaveform}.
The simplest signal distortion technique may be the iterative clipping and filtering (ICF) method
 that reduces PAPR in an intuitive manner.
The clipping procedure directly confines the amplitude of OFDM signal to a preset threshold and then the filtering procedure suppresses the consequent out-of-band emission (OOBE).
Although the classical ICF method has relatively low computational complexity which mainly costs the fast Fourier transform (FFT) and inverse FFT (IFFT) operations,
 multiple repetitions are generally required to achieve the desired PAPR reduction\cite{2002EL-ArmstrongPeak}.
Recently, with the exploitation of optimization method, the distortion based technique has been studied to {\color{black}pursue} the optimal system performance in terms of PAPR and error vector magnitude (EVM).
In \cite{2006ITSP-AggarwalMinimizing}, the PAPR minimization problem with the EVM constraint is formulated as a second order conic programming (SOCP) for the first time.
Subsequently, the SOCP is extended to solve an EVM optimization task subject to a deterministic PAPR constraint and a spectral mask constraint\cite{2009IJSTSP-LiuError}, 
 which is solved by using the interior-point method.
In addition, the filtering procedure of ICF is formulated as an optimization problem to find the optimal filter coefficients in \cite{2011ITC-WangOptimized},
 which is called optimized ICF (OICF) method.
This method is further considered for simplification in \cite{2013ITC-ZhuSimplified}.
Recently, 
the alternating direction method of multipliers (ADMM), as a powerful first-order optimization technique, is widely used in distributed optimization and statistical learning\cite{2010FTML-BoydDistributed}.
As the ADMM is well suited to large-scale convex optimization,
the authors in \cite{2018ITVT-BaoADMM} firstly exploit the ADMM to efficiently solve the PAPR minimization in large-scale multiple-input multiple-output (MIMO) systems.
In \cite{2018ITSP-YongchaoWangOptimized}, the ADMM is introduced to the PAPR optimization problem with theoretical convergence result.

Considering the mixed-numerology transmissions, the existing PAPR reduction techniques cannot be directly applied due to the following reasons.
First, since PAPR is measured for the composite signal before PA, representing the sum of all subbands signals,
the traditional PAPR reduction technique performed on the separate subbands possibly will not level down the PAPR of the composite signal.
Furthermore, unlike other multi-bands OFDM systems,
 such as noncontiguous-OFDM (NC-OFDM) \cite{-Non} and carrier aggregation\cite{2014IN-Z.KhanCarrier},
  in which the subcarriers are still orthogonal to each other and the signal can be modulated by one IFFT/FFT module similar to the conventional OFDM,
 the mixed-numerology systems are equipped with multiple scalable IFFT/FFT modules where the subband signals are generated individually.
How to operate on the separate subbands to achieve the overall PAPR reduction becomes an open question.
To the best of the authors’ knowledge, the mixed-numerology systems are different and lead to the research and development of PAPR reduction techniques that are not currently available in the literature.

\subsection{Contributions}
In this paper, we consider the signal distortion techniques for PAPR reduction in the mixed-numerology systems.
It will be shown that the direct extension of the classical ICF method to the mixed-numerology system results in INI accumulation.
A new noise-shaped ICF (NS-ICF) method is designed to elaborately avoid the increase of INI.
Meanwhile, by formulating the PAPR reduction of composite signals as a multi-block convex program,
a simple reformulation of the constrained convex problem falls in the applicable scope of the ADMM.
As emphasized in \cite{2010FTML-BoydDistributed}, the philosophy of ADMM is a ``decomposition-coordination procedure",
 in which the solutions to small local subproblems are coordinated to find a solution to the large global problem.
Interestingly, the generating process of the composite signal is implicitly similar to the procedure of ADMM.
The PAPR optimization performed on each subband will eventually achieve the global optimized signal. 
We show that the ADMM approach is efficient for the PAPR reduction of composite signal and leads to tractable and scalable algorithms for the cases of any number of numerologies.

The contributions of this paper are summarized as follows:
\begin{itemize}
	\item 
	We first build a system model and analyze the PAPR problem for the OFDM-based mixed-numerology system.
	As the composite signal is composed of multiple individual OFDM signals,
	we illustrate that the high PAPR problem is still troublesome.
	It is also shown that independently applying PAPR reduction technique to separate subbands is incapable of reducing the PAPR of composite signal.
	\item 
	In order to avoid the INI accumulation caused by repeated execution,
	we propose a new clipping-based method, named NS-ICF, as the benchmark for PAPR reduction, which exploits the clipping noise instead of the clipped signal.
	The analytical expression is provided to prove that the NS-ICF method does not introduce extra INI.
	\item 
	We then pursue the optimal performance with respect to the EVM and PAPR.
	A distortion minimization problem with a constraint of PAPR is then formulated as a convex program.	
	Reformulated as a favorable separable structure emerging in both the
	objective function and the constraints, the original problem can be efficiently solved through ADMM approach.
	The proposed ADMM-based algorithms split the original problem into several subproblems whose optimal solution can be determined analytically.
	Moreover, the computational complexity in each iteration is comparable to that of the NS-ICF method.	
	{\color{black} 
	\item	
	The proposed PAPR reduction methods are also considered in combination with the filtering and windowing techniques.
	We show that the idea of NS-ICF can be easily used in filtered-OFDM (F-OFDM) in which the clipping noise is processed by the identical   subband filter in time domain, and the windowing operation is included into the optimization formulation that can be solved by the proposed ADMM-based algorithms.
}
	
\end{itemize}

\subsection{Organization and Notations}
The rest of this paper is organized as follows.
Section II establishes the system model for mixed numerologies transmission and briefly reviews the PAPR problem.
In Section III, we redesign the classical ICF method and apply it to the mixed-numerology system.
Then, the optimization model of PAPR reduction is formulated and two algorithms are proposed in Section IV.
{\color{black}
The applications of proposed PAPR methods with the filtering and windowing techniques are discussed in Section V.}
Section VI presents the simulation results and discussion.
The conclusions are drawn in Section VII.

\textit{Notations:}
In this paper, vectors and matrices are denoted by lowercase and uppercase bold letters, and
 $ (\cdot)^H $ symbolize the Hermitian conjugate operation.
We use $ \|\cdot\|_\infty $ and $ \|\cdot\|_2 $ to denote the $ \infty $-norm and $ 2 $-norm of a vector, respectively.
{\color{black} 
$ {\bf I}_N $ and $ {\bf 0}_{m\times n} $ refer to $ N\times N $ identity matrix and $ m\times n $ zero matrix, respectively.
$ blkdiag({\bf A},N) $ denotes a diagonal matrix generated by $ N $ repetitions of $ {\bf A} $.}
$ \nabla  $ stands for the gradient operator.

\section{Preliminaries}
We start by introducing the system model and then summarize the fundamental PAPR issue arising in the OFDM-based mixed-numerology systems.

\subsection{System Model of Mixed-Numerology Transmission}
We consider a mixed-numerology OFDM system where the system bandwidth is divided into several subbands with different subcarrier spacings.
For a subband with $ K_i $ continuous subcarriers using numerology $ i $,
the continuous-time OFDM symbol can be obtained by
\begin{equation}\label{OFDMsymbol}
\begin{split}
s_{i}^u(t) = {1 \over \sqrt{K_i}}\sum_{k = 0}^{{K_i } - 1}x_{i}^u(k)e^{j2\pi (k f_i+\Delta F_i){\color{black}(t - T_{C\!P,i}-(u-1)T_i)}}&,\\
  uT_i \leq t < (u+1)T_i&,
\end{split}
\end{equation}
where $ x_{i}^u(k) $ is the independent identically distributed complex modulation symbol with zero mean and unit power on the $ k $-th subcarrier of the $ u $-th OFDM symbol.
$ f_i $ denotes the subcarrier spacing of numerology $ i $ and
$ \Delta F_i $ is frequency offset of the $ i $-th subband.
{\color{black}The signal duration is $ T_i = T_{C\!P,i} + T_{sys,i} $, where $ T_{C\!P,i} $ denotes the duration of cyclic prefix (CP) and $ T_{sys,i}=1/f_i $.}

There is a commonly accepted family of numerologies that defines the subcarrier spacings as follows
\cite{2018ICSM-ZaidiOFDM,2017IJoSAiC-Guan5G}
\begin{equation}\label{scs}
 f_i = 2^{v_i} f_{1},  {\color{black} \, T_{C\!P,i}=T_{C\!P,1}/2^{v_i},} \qquad i =1,2,...,M,
\end{equation}
where $ v_i  $ is an integer and $ M $ is the number of OFDM numerologies.
Equation (\ref{scs}) implies that 
 any subcarrier spacing is an integer divisible by all smaller subcarrier spacing.
By this mean, the waveform is scalable in the sense that the subcarrier spacing of OFDM can be chosen according to (\ref{scs}).
For example, in the current version of 5G NR\cite{2017-3GPPNR}, the subcarrier spacing of OFDM can be chosen according to $ 15 \times 2^{v_i} $ kHz where 15 kHz is the subcarrier spacing used in Long Term Evolution (LTE).
With this scaling approach, different OFDM numerologies can be implemented easily by using different-scaled FFT under the same sampling clock rate, which is also illustrated as single rate system\cite{2017ICM-ZhangMulti}.
	
\begin{figure}[t]
	\centering
	\includegraphics[width=80mm]{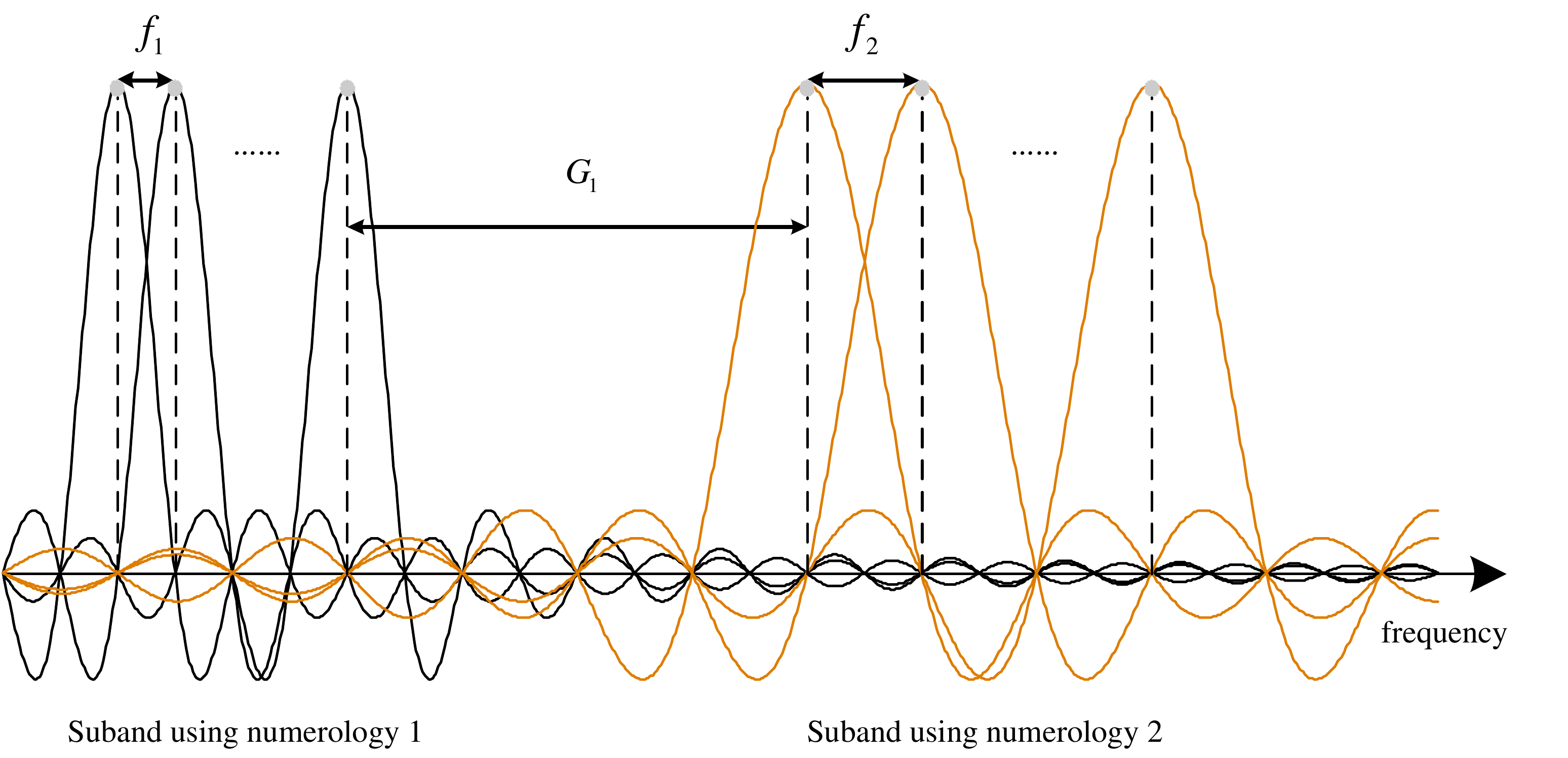}
	\caption{An example of mixed numerologies transmission with two subbands.}
	\label{model1}
\end{figure}

Without loss generality, we assume that $ \Delta F_1=0 $ and the guard band between $ i $-th and $ (i+1) $-th subbands is $ G_{i} $.
Then, the system bandwidth can be roughly expressed as
\begin{equation}\label{}
B=\sum\limits_{i = 1}^{M - 1} {\left( {{K_i}{f_i} + {G_i}} \right)} + {K_M}{f_M}.
\end{equation}

We also use the concept of generalized synchronized system
{\color{black}
whose symbol period is equivalent to the least common multiple (LCM) of duration of all subbands\cite{2017ICM-ZhangMulti}.
It benefits simplified system design
since only limited symbols need to be considered in a processing window, and every LCM symbol has the same overall performance\cite{2017ICM-ZhangMulti}.}
Specifically,
 the signal duration has the relationship of $  T_1 = 2^{v_2} T_2=...=2^{v_M}T_M $.
For example, if $ M=2 $, one LCM symbol is compounded of one OFDM symbol of numerology 1 and two OFDM symbols of numerology 2, i.e.,
\begin{equation}
\begin{split}\label{z}
&\quad z^u\left( t \right) = \\
&\left\{ \!{\begin{array}{*{20}{c}}
	{\eta_1{s_1^u}\left( t \right) + \eta_2{s_{2}^{2u-1}}\left( t \right),}\\
	{\eta_1{s_1^u}\left( t \right) + \eta_2{s_{2}^{2u}}\left( {t - {{{T_1}/{2}}}} \right),}
	\end{array}} \right.\begin{array}{*{20}{c}}
\!\!{uT_1\le t<uT_1+T_2}\\
\!\!{uT_1+T_2} \le t<(u+1)T_1
\end{array}
\end{split}
\end{equation}
where $ z^u\left( t \right) $ is the $ u $-th LCM symbol and
$ \eta_i $ denotes the power adjusting factor for the $ i $-th subband.
Note that if the total transmit power is normalized to $ M $ and evenly distributed on each subband,
 then we have $ {\eta  _i} = 1 $.
{\color{black}Without loss of generality, we focus on the time interval $ \left[ 0,T_1\right)  $ and omit the superscript of $ z\left( t \right) $.
}

\subsection{PAPR Problem}
According to the definition of PAPR which is the ratio of the peak power of the signal to its average power,
we can express the PAPR of the composite signal $ z(t) $ as
\begin{equation}\label{PAPR}
{\rm{PAPR}} = {\max_{t \in [0, T_1)} \left\vert z(t)\right\vert^{2} \over P_{\rm av}},
\end{equation}
where $ P_{av} = E\{|z(t)|^2\} $ is the average power of $ z(t) $.
Nevertheless, analog signals are not amiable for calculation and analysis.
To address this issue, the discrete samples of $ z(t) $ are usually used to compute the PAPR.
To approximate the PAPR of the continuous-time signal accurately,
 $ J $-time oversampling is usually considered\cite{2001ITC-Ochiaidistribution}.
In addition, for the convenience of FFT implementation and the uniform of sampling rate, the $ JN_i $-point IFFT are commonly used to modulate the oversampled OFDM signal, where $ N_i=2^{\lceil log_2(B/f_i)\rceil} $.
Then, the oversampled signal $ s_i^{u_i}(n) $ can be efficiently computed by 
\begin{equation}\label{sn}
\begin{split}
s_i^{u_i}(n) = {1 \over \sqrt{JN_i}}\sum_{k = 0}^{{K_i } - 1}x_i^{u_i}(k)e^{j\frac{{2\pi {\color{black}(n-L_{C\!P,i})}\left( {k + \Delta {k_i}} \right)}}{{J{N_i}}}},\\ \quad 0 \leq n \leq L\!_{sys}-1,
\end{split}
\end{equation}
where $ \Delta {k_i} = {\Delta {F_i}}/{f_i} $, {\color{black}$ L\!_{sys} = JN_i + L_{C\!P,i} $ and $ u_i = 1,2,...,2^{v_i} $}.
In order to facilitate the algebraic operation, we use the matrix representations instead.
The oversampled signal (\ref{sn}) can also be expressed as
\begin{equation}\label{sn2}
{{\bf{s}}_i^{u_i}} = {{\bf{P}}_i}{{\bf{D}}_i}{{\bf{x}}_i^{u_i}},
\end{equation}
where $ {\bf{x}}_{i}^{u_i}=[x_{i}^{u_i}(0),x_{i}^{u_i}(1),...,x_{i}^{u_i}(K_i-1)]^T $.
$ {{\bf{D}}_i} \in {\mathbb{C}^{J{N_i} \times {K_i}}} $ is the first $ K_i $ columns of the $ JN_i $-points normalized and frequency-shifted IFFT matrix which is shifted by $ \Delta {k_i} $,
{\color{black}
and 
$ {{\bf{P}}_i} = [{{\bf{0}}_{L_{C\!P,i}\times(JN_i-L_{C\!P,i})}},{{\bf{I}}_{L_{C\!P,i}}};{{\bf{I}}_{JN_i}}]$ denotes the matrix of CP insertion.
}

Then, 
for an LCM symbol compounded of $ 2^{v_i} $ symbols of numerology $ i $ (for $ i=1,2,...,M $), 
we denote the composite symbol as
\begin{equation}\label{vector z}
{\bf{z}} = {\bf{F}}_1{\bf{x}}_1 + {\bf{F}}_{2}{\bf{x}}_{2} +...+ {\bf{F}}_{M}{\bf{x}}_{M},
\end{equation}
where {\color{black} $ {{\bf{F}}_i} = blkdiag({{\eta _i}{{\bf{P}}_i}{{\bf{D}}_i}},2^{{v_i}}) $
and 
$ {{\bf{x}}_i} = {\left[ {{\bf{x}}_i^1;{\bf{x}}_i^2;...;{\bf{x}}_i^{{2^{{v_i}}}}} \right]} $
for $ i=1,2,...,M $.}

Thus, the PAPR of composite signal $ z(n) $ is given by
\begin{align}\label{papr}
{\rm{PAPR}} = \frac{{\mathop {\max }\limits_{n = 0,1,...,{\color{black}{L\!_{s\!y\!s}}}} {{\left| {z(n)} \right|}^2}}}{{\frac{1}{{\color{black}{L\!_{s\!y\!s}}}}\sum\limits_{n = 0}^{{\color{black}{L\!_{s\!y\!s}}} - 1} {{{\left| {z(n)} \right|}^2}} }} = \frac{{\left\| {\bf{z}} \right\|_\infty ^2}}{{\frac{1}{{\color{black}{L\!_{s\!y\!s}}}}\left\| {\bf{z}} \right\|_2^2}}.
\end{align}

Note that the envelope of the composite signal $ \bf{z} $ is generated from the superposition of all the signals from each subband.
We can infer that the composite signal sampled at the Nyquist rate still converges to a complex Gaussian random process for large $ N_i $.
It can be explained by the fact that the sum of Gaussian distributed random variables obeys Gaussian distribution.
Therefore, the envelope of composite signal becomes a Rayleigh random variable\cite{2001ITC-Ochiaidistribution}, which implies that the high PAPR problem still exists in the mixed-numerology systems and the PAPR reduction techniques should be considered for the composite signal rather than the individual subband signal.

In the following sections, to clearly illustrate our ideas and without loss generality, we only consider the  mixed-numerology transmission with $ M = 2 $ numerologies as shown in Fig. \ref{model1}.
However, the proposed PAPR reduction methods can be straightforwardly extended to the $ M>2 $ cases. 

\section{Iterative Clipping and Filtering method in Mixed-Numerology System}
The main idea of ICF is to hard limit the amplitude of signals beyond the clipping threshold and subsequently eliminate the clipping noise outside the band.
To characterize the fundamental properties, we start from reviewing the classical ICF method in conventional OFDM system.
Then we present the different characteristics of ICF in the mixed-numerology system in full detail and propose a new clipping-based method as a benchmark for PAPR reduction techniques.

\subsection{ICF method}
The clipping and filtering operation are iteratively performed as described in \cite{2002EL-ArmstrongPeak}.
To be specific, the clipping procedure is performed by
\begin{align}\label{clip}
\bar{s}(n) = \left\{ {\begin{array}{*{20}{c}}
	&Ae^{j\theta(n)},  &\left\vert s(n) \right\vert > A\\
	&s(n),   &\left\vert s(n)\right\vert \leq A
	\end{array}} \right.
\end{align}
where $ \bar{s}(n) $ is the clipped signal and $ \theta (n) $ denotes the phase of $ s(n) $.
$ A $ is the clipping level calculated in each iteration according to the predefined clipping ratio (CR),
which is defined as \cite{2013ICST-RahmatallahPeak}
\begin{align}\label{thresh}
{\rm{CR}} = 20\,{\rm{log}}_{10}\left({\dfrac{A}{\frac{1}{{\sqrt {JN} }}{\left\| {\bf{s}} \right\|_2}} }\right) \ ({\text{dB}}).
\end{align}
We can see that CR is calculated by the preset PAPR threshold.
Note that clipping is a nonlinear process that leads to both in-band distortion and OOBE.
In order to eliminate the OOBE, 
 the filtering approach is applied to the clipped signal in frequency domain.
The filter design is based on a rectangular window of which the frequency response is given by \cite{2002EL-ArmstrongPeak}
\begin{equation}\label{filter}
H\left( k \right) = \left\{ {\begin{array}{*{20}{c}}
	{1,}\\
	{0,}
	\end{array}} \right.\begin{array}{*{20}{c}}
{1 \le k \le N}\\
{N + 1 \le k \le JN}
\end{array}.
\end{equation}

What should be further illustrated is that the clipping noise is defined as the difference between the clipped and original signals, which is expressed as\cite{2000IJSAC-OchiaiPerformance}
\begin{equation}\label{cn}
d\left( n \right) = \bar s\left( n \right) - s\left( n \right) = \left\{ {\begin{array}{*{20}{c}}
	{\left( {A - \left| {s\left( n \right)} \right|} \right){e^{j\phi \left( n \right)}},}\\
	{0,}
	\end{array}} \right.\!\!\begin{array}{*{20}{c}}
{\left| {s\left( n \right)} \right| > A}\\
{\left| {s\left( n \right)} \right| \le A}
\end{array}\!.
\end{equation}

Additionally, the power spectrum density (PSD) of the clipping noise is evenly distributed over the system band\cite{2008ITVT-WangAnalysis} where the in-band power distorts the transmitted symbol and the out-of-band power leads to low spectrum efficiency.
Although the filtering approach can eliminate the OOBE, the  peak regrowth would occur as a side effect.
Therefore, repeated clipping and filtering are always required to achieve the desired PAPR.

\subsection{ICF method for Mixed-Numerology Transmitted Signal}
\begin{figure*}[htbp]
	\centering
	\subfigure[The classical ICF method]{
		\begin{minipage}{0.99\linewidth}	
			\centering		
			\includegraphics[width=160mm]{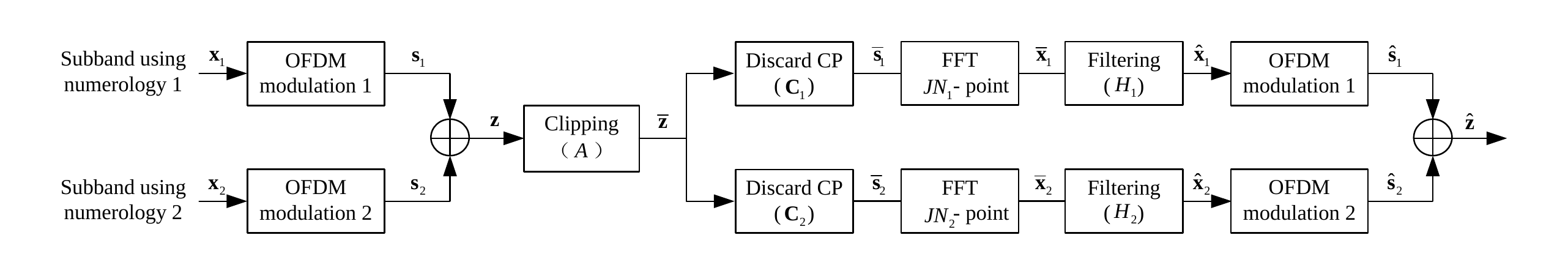}
			\label{fig2_1}
		\end{minipage}
	}
\\
	\subfigure[The propoed NS-ICF method]{
		\begin{minipage}{0.99\linewidth}
			\centering	
			\includegraphics[width=160mm]{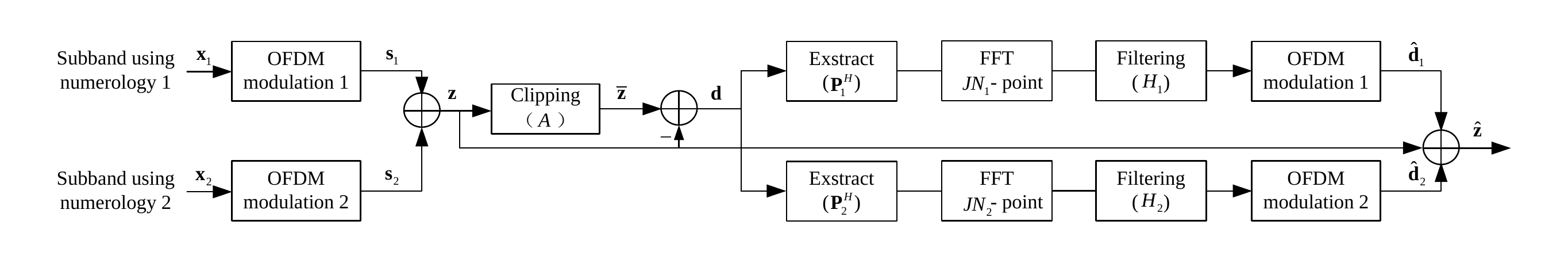}
			\label{fig2_2}
		\end{minipage}
	}	  
	\caption{{\color{black}The diagrams of the ICF and NS-ICF methods.}}
	\label{fig2}
\end{figure*}

As shown in Fig. 2(a), we can readily extend the ICF method to the mixed-numerology system by clipping the composite signal $ \bf{z} $ and filtering the clipped version in individual subbands, 
{\color{black}where $ \bar{\bf{s}}_i $ is obtained by removing the beginning $ L_{C\!P,i} $ samples of $ \bar{\bf z} $ with the matrix $ {\bf{C}}_i = [{\bf{0}}_{J\!N_i\times L_{C\!P,i}},{\bf{I}}_{J\!N_i}]$.}
Then, by using $ JN_i $-point FFT to transform $ \bar{\bf s}_i $ to the frequency domain, we have the clipped symbol $ \bar{\bf x}_i $.
The following filtering $ {\bf H}_i=diag(H_i(1),...,H_i(JN_i)) $ is applied to $ \bar{\bf x}_i $, from which the components located in each subband are extracted and the out-of-band components are set to zero.
After frequency filtering, $ \hat{\bf x}_1 $ and $ \hat{\bf x}_2 $ are modulated to OFDM symbols that are then added together to be transmitted or put into the next execution.

However, the output symbol of the filter contains not only the distorted original symbol but the interference from the other subband. 
For example, $ \hat{\bf x}_1 $ can be expressed as
{\color{black}
\begin{equation}
\begin{split}\label{filteredx}
{{\bf{\hat x}}_1} &= {\bf{D}}_1^H {{\bf{\bar s}}_1} = {\bf{D}}_1^H {\bf{C}}_1({{\bf{z}} + {\bf{d}}}) 
\\&= {{\bf{x}}_1} + \underbrace{{\bf{D}}_1^H{\bf{C}}_1{{\bf{F}}_2}{{\bf{x}}_2}}_{\text{INI}}
+\underbrace{{\bf{D}}_1^H{\bf{C}}_1{\bf{d}}}_{\text{in-band distortion}},
\end{split}
\end{equation}
where the FFT operation is represented by $ {\bf{D}}_1^H $.
As $ {\bf{D}}_1 $ only contains $ K_1 $ columns of normalized IFFT matrix to modulate the used subcarriers, filtering operation $ {\bf H}_1 $ can be omitted in (\ref{filteredx}).}
We can observe that the original symbol is distorted by the in-band clipping noise which is denoted as the third term in the left of the equation from (\ref{filteredx}).
The second term is the INI introduced by the subband using numerology 2.
Since the orthogonality is no longer held for the waveforms using different numerologies,
the leakage energy will be caught by the other subband.
The output composite signal is then given by
{\color{black}
\begin{equation}\label{z_clip}
\begin{split}
\hat{\bf{z}}=&{{\bf{F}}_1}{{{\bf{\hat x}}}_1} + {{\bf{F}}_2}{{{\bf{\hat x}}}_2} 
\\
= &{\bf{z}} 
+ \underbrace{ {{\bf{F}}_1}{\bf{D}}_1^H{\bf{C}}_1{{\bf{F}}_2}{{\bf{x}}_2} + {{\bf{F}}_2}{\bf{D}}_2^H{\bf{C}}_2{{\bf{F}}_1}{{\bf{x}}_1}}_{\text{INI}}\\
& \qquad\qquad +
\underbrace{\left({\bf{F}}_1{\bf{D}}_1^H{\bf{C}}_1 + {\bf{F}}_2{\bf{D}}_2^H{\bf{C}}_2 \right){\bf{d}}}_{\text{in-band distortion}} 
\end{split}.
\end{equation}

Although the INI can be reduced by broadening the guard band,
the repetition progress will inevitably accumulate the INI that remains in the output signal at each execution. 
The interference level is increasingly reinforced as the ICF method is repeatedly executed.
In addition, the high amplitudes at the beginning $ L_{C\!P,i} $ samples of the signal period is not considered after the CP removal, so that high peaks are probably still in existence.}
Therefore, the straightforward application of the classical ICF method to the mixed-numerology systems seems infeasible.

In order to avoid INI and inherit the idea of the ICF method, we then focus on the clipping noise.
{\color{black}
As shown in Fig. \ref{fig2_2}, the clipping noise rather than the clipped signal are used for frequency-domain filtering in the individual subbands.
Moreover, the beginning $ L_{C\!P,i} $ samples of the clipping noise are added to the end of the signal using matrix $ {\bf P}_i^H $.

The filtered clipping noise is then written as
\begin{equation}\label{cn_f}
{{\bf{\hat d}}_i} ={{\bf{F}}_i}{\bf{D}}_i^H{\bf{P}}_i^H{\bf{d}} ={{\bf{F}}_i}{\bf{F}}_i^H{\bf{d}},\qquad i=1,2.
\end{equation}}
By this mean, the output composite signal can be obtained by
\begin{equation}\label{z_clip2}
{\bf{\hat z}} = {\bf{z}} + {{\bf{\hat d}}_1} + {{\bf{\hat d}}_2} = {\bf{z}} + \left( {{{\bf{F}}_1}{\bf{F}}_1^H + {{\bf{F}}_2}{\bf{F}}_2^H} \right){\bf{d}}.
\end{equation}
Obviously, the terms of INI no longer appear in (\ref{z_clip2}) compared with (\ref{z_clip}).
On account that the filtering is applied to the clipping noise, we call it the noise-shaped ICF method.
Interestingly, while the NS-ICF method is equivalent to the classical ICF method in the conventional single numerology cases, the classical ICF method has to be redesigned as the NS-ICF method for the mixed-numerology systems.


Note that the NS-ICF method is very simple and straightforward and can serve as a benchmark for the distortion based PAPR reduction techniques.
Due to the peak regrowth, it generally requires many times of execution to achieve the desired PAPR.
In addition, the distortion caused by amplitude clipping is not considered.
In the next section, the optimization method is proposed to achieve both the PAPR reduction and distortion control.

\section{PAPR reduction using Optimization Method}

In this section, we first look back the OICF method and then establish the optimization problem for the PAPR reduction of composite signal.
As a common approach to solve optimization problems in the form of SOCP, interior-point method has been applied for the PAPR reduction problem\cite{2009IJSTSP-LiuError,2011ITC-WangOptimized}.
However, prohibitively high computation complexity is often required for the system with large number of subcarrier.
Thus, algorithm efficiency is of extreme importance for practical application.
By reformulating the optimization problem as a linearly constrained separable convex programming, 
whose objective function is separable into multiple individual convex functions without coupled variables, 
we show that the resulting model falls into the applicable scope of the well-known ADMM approach and can be efficiently solved by splitting it into some more tractable subproblems.

\subsection{OICF method}
The OICF method aims to minimize the in-band distortion under the PAPR constraint.
The symbol-wise distortion can be quantified by the EVM, which is defined as the square root of the ratio of the mean error vector power to the mean reference power, i.e.,
\begin{equation}\label{EVM}
{\rm{EVM}}_{\bf{x}} = \sqrt {\frac{{{\textstyle{1 \over N}}\sum\nolimits_{k = 0}^{N - 1} {{{\left| {{x(k)} - {{\hat x}(k)}} \right|}^2}} }}{{{\textstyle{1 \over N}}\sum\nolimits_{k = 0}^{N - 1} {{{\left| {{x(k)}} \right|}^2}} }}}  = \frac{{{{\left\| {\bf{x} - \hat {\bf{x}}} \right\|}_2}}}{{{{\left\| \bf{x} \right\|}_2}}},
\end{equation}
where $  {\bf{x}} $ is the original OFDM symbol and $ {\hat {\bf{x}}} $ denotes the modified symbol generated by PAPR reduction technique.
Then the OICF method can be formulated as \cite{2011ITC-WangOptimized}
\begin{subequations}
	\label{OICF}
	\begin{align}
	\mathop {{\rm{min}}}\limits_{\hat{\bf{x}} \in {\mathbb{C}^N}} & \quad \frac{{{{\left\| {\bf{x} - \hat {\bf{x}}} \right\|}_2}}}{{{{\left\| \bf{x}\right\|}_2}}},\\
	s.t. \quad
	&{{\hat {\bf{s}}} } = {\rm{IFFT}}{( {\hat{\bf{x}} } )}
	\label{OICF,b},\\
	&\frac{{{{\left\| {{{\hat {\bf{s}}}}} \right\|}_\infty }}}{{{{{{\left\| {{{\hat {\bf{s}}} }} \right\|}_2}} / {\sqrt {L\!_{s\!y\!s}} }}}}  \le \sqrt {{\rm{CR}}}  = \gamma.
	\label{OICF,c}
	\end{align}
\end{subequations}
Note that the problem (\ref{OICF}) considers the optimized OFDM symbol instead of filter coefficients as the optimization variable,
which is an equivalent form to determine the optimal filter design\cite{2013ITC-ZhuSimplified}.
To relax the non-convex constraint (\ref{OICF,c}) to its convex counterpart that makes the problem solvable,
$ \left\|{\hat {\bf{s}}}  \right\| $ is replaced with the original version $ \|{ {\bf{s}}} \| $.
Then, the constraint (\ref{OICF,c}) is rewritten as
\begin{equation}\label{appr}
{\| {\hat {\bf{s}}} \|}_\infty   \le \gamma \|{ {\bf{s}}} \|_2/\sqrt{L\!_{s\!y\!s}}.
\end{equation}
With this modification, the problem (\ref{OICF}) is formulated as an SOCP, to which many well-known optimization algorithms such as interior-point method can be applied.

\subsection{Optimization Formulation}
We now consider the distortion in an LCM symbol.
Referring to the conception of EVM (\ref{EVM}), we define the EVM of a composite signal with $ M $ different numerologies as the square root of total normalized distortion power on every sub-symbol,
which is
\begin{equation}\label{}
{\rm{EV}}{{\rm{M}}_{\bf{z}}} = \sqrt {\sum\limits_{i = 1}^M {\frac{1}{{{2^{{v_i}}}}}\left( {\sum_{v = 1}^{{2^{{v_i}}}} {{\rm{EVM}}_{{\bf{x}}_i^v}^2} } \right)} }.
\end{equation}
In the case that only the first and second numerologies are used,
the EVM of the composite signal $ \bf z $ can be written as
\begin{equation}\label{EVMz}
\begin{split}
{\rm{EVM}}_{\bf{z}} = \sqrt {\frac{{\left\| {{{\bf{x}}_1} - {{{\bf{\hat x}}}_1}} \right\|_2^2}}{{\left\| {{{\bf{x}}_1}} \right\|_2^2}} + \frac{{\left\| {{{\bf{x}}_2} - {{{\bf{\hat x}}}_2}} \right\|_2^2}}{{\left\| {{{\bf{x}}_2}} \right\|_2^2}}} .
\end{split}
\end{equation}

To minimize the distortion of an LCM symbol and constrain the PAPR,
we formulate the following optimization problem
\begin{subequations}
	\label{Op}
	\begin{align}\label{Op,a}
	\mathop {{\rm{min}}}\limits_{{{\hat {\bf{x}}}_1},{{\hat {\bf{x}}}_{2}}}\quad
	&\sqrt {\frac{{\left\| {{{\bf{x}}_1} - {{{\bf{\hat x}}}_1}} \right\|_2^2}}{{\left\| {{{\bf{x}}_1}} \right\|_2^2}} + \frac{{\left\| {{{\bf{x}}_2} - {{{\bf{\hat x}}}_2}} \right\|_2^2}}{{\left\| {{{\bf{x}}_2}} \right\|_2^2}}},
	\\
	\label{Op,b}
	{s.t.}\quad 
	&  {\frac{{\left\| {{\bf{F}}_1{\hat {\bf{x}}}_1 + {\bf{F}}_{2}{\hat {\bf{x}}}_{2}} \right\|_\infty }}{{\frac{1}{\sqrt{L\!_{s\!y\!s}}}\left\| {{\bf{F}}_1{\hat {\bf{x}}}_1 + {\bf{F}}_{2}{\hat {\bf{x}}}_{2}} \right\|_2}} \le \gamma },
	\end{align}
\end{subequations}
where $ {{\hat {\bf{x}}}_1} \in {\mathbb{C}^{{K_1}}},{{\hat {\bf{x}}}_{2}} \in {\mathbb{C}^{{K_1}}} $.
Similar to (\ref{OICF}), the constraint (\ref{Op,b}) is also a non-convex inequity that makes numerical solution of (\ref{Op}) difficult.
Following the same procedure as mentioned in (\ref{appr}), 
$ \frac{1}{\sqrt{J{N_1}}}{\left\| {{\bf{F}}_1{\hat {\bf{x}}}_1 + {\bf{F}}_{2}{\hat {\bf{x}}}_{2}} \right\|_2} $ can be approximated by  $ \frac{1}{\sqrt{J{N_1}}}{\left\| {{\bf{z}}} \right\|_2} $.
Then the problem (\ref{Op}) is rewritten in the following form
\begin{subequations}
\label{originOp}
\begin{align}
\mathop {{\rm{min}}}\limits_{{{\hat {\bf{x}}}_1},{{\hat {\bf{x}}}_{2}}}\quad
&\sqrt {\frac{{\left\| {{{\bf{x}}_1} - {{{\bf{\hat x}}}_1}} \right\|_2^2}}{{\left\| {{{\bf{x}}_1}} \right\|_2^2}} + \frac{{\left\| {{{\bf{x}}_2} - {{{\bf{\hat x}}}_2}} \right\|_2^2}}{{\left\| {{{\bf{x}}_2}} \right\|_2^2}}},
\\
\label{originOp,b}
{s.t.}\quad 
&  {{{\left\| {{\bf{F}}_1{\hat {\bf{x}}}_1 + {\bf{F}}_{2}{\hat {\bf{x}}}_{2}} \right\|_\infty }} \le \gamma\frac{1}{\sqrt{L\!_{s\!y\!s}}}{\left\| {{\bf{z}}} \right\|_2} },
\end{align}
\end{subequations}
which is also an SOCP that can be solved by resorting to some open source software like CVX\cite{CVX}.

In order to find more efficient algorithm, we continue with this optimization problem (\ref{originOp}).
Since the problem $ \mathop {\min }\limits_{\bf{x}} {\left\| {{\bf{Ax}} - {\bf{b}}}  \right\|_2} $ has the same optimal solution as $ \mathop {\min }\limits_{\bf{x}} {\left\| {{\bf{Ax}} - {\bf{b}}}  \right\|_2^2} $\cite{2004-BoydConvex},
we can square the objective function and then obtain a sum of two separable convex functions.
Moreover, an auxiliary variable $ {\hat{\bf z}}\in{\mathbb{C}}^{L\!_{s\!y\!s}} $ can be introduced to the optimization problem, which has the linear relationship with $ {\hat{\bf x}}_1, {\hat{\bf x}}_2 $, i.e.,
$ {\hat {\bf{z}}} = {\bf{F}}_1{\hat {\bf{x}}}_1 + {\bf{F}}_{2}{\hat {\bf{x}}}_{2} $.
Taking these aspects into consideration, the original problem can be then reformulated as
\begin{subequations}
	\label{Op2}
	\begin{align}\label{Op2,a}
	\mathop {{\rm{min}}}\limits_{{{\hat {\bf{x}}}_1},{{\hat {\bf{x}}}_{2}},\hat{\bf{z}}} \quad
	&\frac{1}{2\sigma_1^2}{{\left\| {{{\bf{x}}_1} - {{{\bf{\hat x}}}_1}} \right\|_2^2}} + \frac{1}{2\sigma_2^2}{{\left\| {{{\bf{x}}_2} - {{{\bf{\hat x}}}_2}} \right\|_2^2}}\\
	\label{Op2,b}
	{s.t.} \quad
	&\hat{\bf{z}} = {{\bf{F}}_1{\hat {\bf{x}}}_1 + {\bf{F}}_{2}{\hat {\bf{x}}}_{2}},
	 \\
	\label{Op2,c}
	& {{\left\| \hat{\bf{z}} \right\|_\infty }} \le \gamma \frac{1}{\sqrt{L\!_{s\!y\!s}}}{\left\| {{\bf{z}}} \right\|_2},
	\end{align}
\end{subequations}
where $ \sigma_i^2= {\left\| {{{\bf{ x}}}_i} \right\|_2^2}$ for $ i=1,2 $.
Note that the optimization objective (\ref{Op2,a}) is the sum of two separate convex functions and the variables obey a linear constraint (\ref{Op2,b}).
This separable structure in both the objective function and constraints in
(\ref{Op2}) enables us to partition the original problem in the light of the ADMM.
By this mean, the minimization problem of (\ref{Op2,a}) can be decomposed into three smaller ones which solve the variables $ {\bf{\hat x}}_1 $, $ {\bf{\hat x}}_2 $ and $ {\bf{\hat z}} $ separably in the consecutive order.

\subsection{O-ADMM Algorithm Framework}
The basic idea of the ADMM is to search for a saddle point of the augmented Lagrangian function rather than directly solve the original constrained optimization problem.
We first give the augmented Lagrangian function of the problem (\ref{Op2})
\begin{equation}\label{Lag}
\begin{split}
{L_\rho }({{\bf{\hat x}}_1},{{\bf{\hat x}}_2},{\bf{\hat z}},{\bf{y}}) = 
\frac{1}{2\sigma_1^2}{{\left\| {{{\bf{x}}_1} - {{{\bf{\hat x}}}_1}} \right\|_2^2}}  +& \frac{1}{2\sigma_2^2}{{\left\| {{{\bf{x}}_2} - {{{\bf{\hat x}}}_2}} \right\|_2^2}} \\
+ \left\langle {{\bf{y}},{{\bf{F}}_1}{{{\bf{\hat x}}}_1} + {{\bf{F}}_2}{{{\bf{\hat x}}}_2} - {\bf{\hat z}}} \right\rangle 
 + \frac{\rho }{2}&\left\| {{{\bf{F}}_1}{{{\bf{\hat x}}}_1} + {{\bf{F}}_2}{{{\bf{\hat x}}}_2} - {\bf{\hat z}}} \right\|_2^2,
\end{split}
\end{equation}
where $ \rho > 0 $ is the penalty parameter, 
$ {\bf{y}} \in \mathbb{C}^{L\!_{s\!y\!s}}$ is the Lagrangian multiplier,
 and the inner product is defined as 
 $\left\langle {{\bf{a}},{\bf{b}}} \right\rangle  \buildrel\textstyle.\over= {\mathop{\rm Re}\nolimits} \left\{ {{{\bf{a}}^H}{\bf{b}}} \right\}$.
The proposed O-ADMM algorithm carries out the following iterative steps,
\begin{subequations}
	\label{iterate}
	\begin{align}\label{iterate,a}
	\hat{\bf{x}}_1^{l + 1} & = \mathop {\arg \min }\limits_{{{\bf{x}}_1} } {L_\rho }\left( {{\hat{\bf{x}}_1},\hat{\bf{x}}_{2}^l,{\hat{\bf{z}}^l},{{\bf{y}}^l}} \right),\\
	\label{iterate,b}
	\hat{\bf{x}}_{2}^{l + 1} & = \mathop {\arg \min }\limits_{{\hat{\bf{x}}_{2}}} {L_\rho }\left( {\hat{\bf{x}}_1^{l + 1},\hat{\bf{x}}_{2},{\hat{\bf{z}}^l},{{\bf{y}}^l}} \right),\\
	\label{iterate,c}
	{\hat{\bf{z}}^{l + 1}} & = \mathop {\arg \min }\limits_{{\bf{\hat z}} } {L_\rho }\left( {\hat{\bf{x}}_1^{l + 1},\hat{\bf{x}}_{2}^{l + 1},\hat{\bf{z}},{{\bf{y}}^l}} \right),\\
	\label{iterate,d}
	{{\bf{y}}^{l + 1}} & = {{\bf{y}}^l} + \rho \left( { {\bf{F}}_1{\hat {\bf{x}}}_1^{l + 1} + {\bf{F}}_{2}{\hat {\bf{x}}^{l + 1}}_{2}  - {\hat{\bf{z}}^{l + 1}}} \right),
	\end{align}
\end{subequations}
where $ l $ is the iteration number.
In each iteration, the O-ADMM algorithm consists of $ \hat{\bf{x}}_1 $-minimization step, $ \hat{\bf{x}}_2 $-minimization step, $ \hat{\bf{z}} $-minimization step, and dual variable update.
Although each step of (\ref{iterate,a}), (\ref{iterate,b}) and (\ref{iterate,c}) needs to solve an optimization problem, we will show that each has a simple closed-form solution and can be solved efficiently.

Based on the augmented Lagrangian function $ {L_\rho }({\hat{\bf{x}}_1},{\hat{\bf{x}}_2},\hat{\bf{z}},{\bf{y}})  $, the problem (\ref{iterate,a}) becomes equivalent to
\begin{equation}
	\label{suba}
	\mathop {\min }\limits_{{\hat{\bf{x}}_1}} 
	\frac{1}{2\sigma_1^2}{{\left\| {{{\bf{x}}_1} - {{{\bf{\hat x}}}_1}} \right\|_2^2}} + \frac{\rho }{2}\left\| {{{\bf{F}}_1}\hat{\bf{x}}_1 + {{\bf{F}}_{2}}\hat{\bf{x}}_{2}^l - {\hat{\bf{z}}^l} + \frac{{{\bf{y}}^l}}{\rho }} \right\|_2^2.
\end{equation}
By setting the derivative of the augmented Lagrangian $ {L_\rho }\left( {{\hat{\bf{x}}_1},\hat{\bf{x}}_{2}^l,{\hat{\bf{z}}^l},{{\bf{y}}^l}} \right) $ to zero, we can obtain the closed-form solution of $ \hat{\bf{x}}_1 $-minimization as
\begin{equation}\label{xk+1}
{\bf{\hat x}}_1^{l + 1} =  \left( {{\frac{1}{\sigma_1^2}{\bf I}_{K_1} + {\eta_1^2\rho{\bf F}_1^H{\bf F}_1 }}}\right)^{-1}\left( {\frac{1}{\sigma_1^2}{{{\bf{x}}}_1} - {\bf{v}}_1^{l}} \right),
\end{equation}
where 
\begin{equation}\label{}
{\bf{v}}_1^{l} = \rho {\bf{F}}_1^H\left( {{{\bf{F}}_{2}}{\bf{\hat x}}_{2}^l - {{\bf{\hat z}}^l} + \frac{{{y^l}}}{\rho }}\right).
\end{equation}


Similarly, the $ \hat{\bf{x}}_2 $-minimization step can be solved by 
\begin{equation}\label{x2}
{\bf{\hat x}}_2^{l + 1} = \left( {{\frac{1}{\sigma_2^2}{\bf I}_{K_1} + {\eta_2^2\rho{\bf F}_2^H{\bf F}_2 }}}\right)^{-1} \left(  \frac{1}{\sigma_2^2}{{{\bf{x}}}_2} -{\bf{v}}_2^l  \right),
\end{equation}
where
\begin{equation}\label{v2}
	{\bf{v}}_2^l =\rho {\bf{F}}_2^H\left( {{{\bf{F}}_1}{\bf{\hat x}}_1^{l + 1} - {{\bf{\hat z}}^l} + \frac{{{{\bf{y}}^l}}}{\rho }} \right).
\end{equation}

For the $ \hat{\bf{z}} $-minimization step,
 we first denote 
\begin{equation}\label{u}
{{\bf{u}}^l}={{\bf{F}}_1}\hat{\bf{x}}_1^{l + 1} + {{\bf{F}}_2}\hat{\bf{x}}_2^{l + 1}  + {{\frac{1}{\rho}{{{\bf{y}}^l}} }}.
\end{equation}
Then, the subproblem (\ref{iterate,d}) is equivalent to
\begin{subequations}
\label{subc}
\begin{align}\label{subd,a}
\mathop {\min }\limits_{{\hat{\bf{z}}} \in \mathbb{C}^{JN_1}} 
& \left\| {{{\bf{u}}^l} - \hat{\bf{z}}} \right\|_2^2,\\
s.t. \quad& {{\left\| \hat{\bf{z}} \right\|_\infty ^2}} \le \gamma \frac{1}{\sqrt{L\!_{s\!y\!s}}}{\left\| {{\bf{z}}} \right\|_2}.
\label{subd,b}
\end{align}
\end{subequations}
Clearly, the problem (\ref{subc}) searches for a vector $ \hat {\bf{z}} $ that is nearest to the vector $ \bf{u} $ and lies within the PAPR threshold,
which can be split into $ L\!_{s\!y\!s} $ one-dimension subproblems as
\begin{subequations}
	\label{zupdate}
	\begin{align}\label{}
	\mathop {\min }\limits_{{{{\hat{z}^{l + 1}}(n)}} \in \mathbb{C}} 
	& |{{{{u}^l(n)}} - \hat{{z}}(n)}|^2 ,\\
	s.t. \quad& {{ \hat{z}(n) }} \le \gamma \frac{1}{\sqrt{L\!_{s\!y\!s}}}{| {{z}}(n)|_2}.
	\label{}
	\end{align}
\end{subequations}

We can find that the optimal solution of (\ref{zupdate}) is a clipping procedure of vector $ {\bf u}^l $.
Hence the $ \hat{\bf{z}} $-minimization step can be easily obtained by
\begin{align}\label{clip_u}
{\hat{z}}^{l+1}(n) = \left\{ {\begin{array}{*{20}{c}}
	&Ae^{j\angle u^l(n)},  &\left\vert u^l(n) \right\vert > A\\
	&u^l(n),   &\left\vert u^l(n)\right\vert \leq A
	\end{array}} \right.,
\end{align}
where $ A = \gamma{\left\| {{\bf{z}}} \right\|_2}/{\sqrt{L\!_{s\!y\!s}}} $.
The proposed ADMM-based algorithm is detailed in Algorithm 1 and referred to as the origin-ADMM algorithm.

\begin{algorithm}[t]
	\caption{\textbf{}O-ADMM algorithm}
	\begin{algorithmic}[1]
		\REQUIRE ~~\\	
		Set $ k=0 $ and initialize ($ \hat{\bf{x}}_1^0,\hat{\bf{x}}_2^0,\hat{\bf{z}}^0,\hat{\bf{y}}^0$).\\
		Select the parameters ($ \rho,\gamma$).\\
		Calculate $ A = \gamma{\left\| {{\bf{z}}} \right\|_2}/{\sqrt{L\!_{s\!y\!s}}} $.
		\ENSURE
		\STATE \textbf{Repeat} $ l $ \\
		\STATE
		Process the $ \hat{\bf{x}}_1 $-minimization step.\\
		(1) Calculate $ {\bf{v}}_1^{l} = \rho {\bf{F}}_1^H\left( {{{\bf{F}}_{2}}{\bf{\hat x}}_{2}^l - {{\bf{\hat z}}^l} + \frac{{{y^l}}}{\rho }}\right) $.\\
		(2) Calculate \\$ {\bf{\hat x}}_1^{l + 1} =  \left( {{\frac{1}{\sigma_1^2}{\bf I}_{K_1} + {\eta_1^2\rho{\bf F}_1^H{\bf F}_1 }}}\right)^{-1}\left( {\frac{1}{\sigma_1^2}{{{\bf{x}}}_1} - {\bf{v}}_1^{l}} \right) $.
		\STATE
		Process the $ \hat{\bf{x}}_2 $-minimization step.\\
		(1) Calculate $ {\bf{v}}_2^l =\rho {\bf{F}}_2^H\left( {{{\bf{F}}_1}{\bf{\hat x}}_1^{l + 1} - {{\bf{\hat z}}^l} + \frac{{{{\bf{y}}^l}}}{\rho }} \right) $.\\
		(2) Calculate \\$ {\bf{\hat x}}_2^{l + 1} = \left( {{\frac{1}{\sigma_2^2}{\bf I}_{K_1} + {\eta_2^2\rho{\bf F}_2^H{\bf F}_2 }}}\right)^{-1} \left(  \frac{1}{\sigma_2^2}{{{\bf{x}}}_2} -{\bf{v}}_2^l  \right) $.		
		\STATE
		Process the $ \hat{\bf{z}} $-minimization step.\\
		(1) Calculate $ {{\bf{u}}^l}={{\bf{F}}_1}\hat{\bf{x}}_1^{l + 1} + {{\bf{F}}_2}\hat{\bf{x}}_2^{l + 1}  + {{\frac{1}{\rho}{{{\bf{y}}^l}} }}  $.\\
		(2) Calculate \\$ 
		{\hat{z}}^{l+1}(n) = \left\{ {\begin{array}{*{20}{c}}
			&Ae^{j\angle u^l(n)},  &\left\vert u^l(n) \right\vert > A\\
			&u^l(n),   &\left\vert u^l(n)\right\vert \leq A
			\end{array}} \right.
	$.
		\STATE
		Update the Lagrangian dual variable $ \hat{\bf{y}} $.\\
		Calculate $ {{\bf{y}}^{l + 1}}  = {{\bf{y}}^l} + \rho \left( { {\bf{F}}_1{\hat {\bf{x}}}_1^{l + 1} + {\bf{F}}_{2}{\hat {\bf{x}}^{l + 1}}_{2}  - {\hat{\bf{z}}^{l + 1}}} \right) $. 
		\STATE
		\textbf{until} reach the stop-criterion.
		Then output $ \hat{\bf{z}}^l$.
	\end{algorithmic}
	\label{Algorithm}
\end{algorithm}

\subsection{CU-ADMM Algorithm Framework}
Due to the convex approximation of the constraint (\ref{Op,b}), the O-ADMM algorithm is probably unable to reach the PAPR threshold after executing once. 
Several repetitions are generally needed to make the constraint (\ref{originOp,b}) approximate to (\ref{Op,b}).
In this subsection, inspired by the projection operation in the $ \hat{\bf{z}} $-minimization step, we propose an improved algorithm in which the constraint of (\ref{subd,b}) is updated in each iteration and the rest of steps in O-ADMM are reused.

Instead of using the original composite signal $ {\bf{z}} $ to calculate the clipping level $ A $ in (\ref{clip_u}),
we exploit the $ \hat{\bf{z}}^l $ to update $ A^l $ before the $ \hat{\bf{z}} $-minimization step, which is written by
\begin{equation}\label{Ak}
A^{l+1} = \gamma \frac{1}{{\sqrt{L\!_{s\!y\!s}}}}{\| {\hat{\bf{z}}^{l}}\|_2}.
\end{equation}
With this modification, the constraint (\ref{Op2,c}) is set according to the instantaneous composite signal $ \hat{\bf{z}}^{l} $,
which implies that the original non-convex constraint (\ref{Op,b}) is approximated by a series of iteratively updated convex constraints. 
Therefore, the $ \hat{\bf{z}} $-minimization step can be rewritten as
\begin{align}\label{clip_u2}
{\hat{z}}^{l+1}(n) = \left\{ {\begin{array}{*{20}{c}}
	&A^{l+1}e^{j\angle u^{l}(n)},  &\left\vert u^{l}(n) \right\vert > A^{l+1}\\
	&u^{l}(n),   &\left\vert u^{l}(n)\right\vert \leq A^{l+1}
	\end{array}} \right.,
\end{align}
where $ A^{l+1} $ is calculated as (\ref{Ak}).
{\color{black}
We can see that the feasible region (\ref{Op2,c}) will gradually shrink as the iteration proceeds and the final optimized output $ \hat{\bf{z}} $ will closely satisfy the PAPR constraint when the algorithm reaches the stop criterion.
}
The proposed algorithm is referred as the constraint update-ADMM and summarized in Algorithm 2.


\begin{algorithm}[t]
	\label{algorithm2}
	\caption{\textbf{}CU-ADMM algorithm}
	\begin{algorithmic}[1]
		\REQUIRE ~~\\	
		Set $ k=0 $ and initialize ($ \hat{\bf{x}}_1^0,\hat{\bf{x}}_2^0,\hat{\bf{z}}^0,\hat{\bf{y}}^0$).\\
		Select the parameters ($ \rho,\gamma$).
		\ENSURE
		\STATE \textbf{Repeat} $ l $ \\
		\STATE
		Update the constraint.\\
		Calculate $ A^{l+1} = \gamma{\left\| {\hat{\bf{z}}^{l}} \right\|_2}/{\sqrt{L\!_{s\!y\!s}}} $.
		\STATE
		Process the $ \hat{\bf{x}}_1 $-minimization step like O-ADMM algorithm.\\		
		\STATE
		Process the $ \hat{\bf{x}}_2 $-minimization step like O-ADMM algorithm.\\
		\STATE
		Process the $ \hat{\bf{z}} $-minimization step.\\
		${\hat{z}}^{l+1}(n) = \left\{ {\begin{array}{*{20}{c}}
			&A^{l+1}e^{j\angle u^{l}(n)},  &\left\vert u^{l}(n) \right\vert > A^{l+1}\\
			&u^{l}(n),   &\left\vert u^{l}(n)\right\vert \leq A^{l+1}
			\end{array}} \right.$		
		\STATE
		Update the Lagrangian dual variable $ \hat{\bf{y}} $ like O-ADMM algorithm.\\		
		\STATE
		\textbf{until} reach the stop-criterion.
		Then output $ \hat{\bf{z}}^l$.
	\end{algorithmic}
	\label{Algorithm2}
\end{algorithm}

\subsection{Computational Complexity}
For the proposed O-ADMM algorithm,
the computation complexity of calculating $ A $ in the initialization stage is $ \mathcal{O}(L\!_{s\!y\!s}) $.
In the $ \hat{\bf{x}}_1 $-minimization step, the main complexity depends on the calculation of $ {\bf{v}}_1^l $, in which the matrix multiplication is equivalent to a $ JN_1 $-FFT operation.
The computation complexity of $ JN_1 $-point FFT is $ \mathcal{O}(JN_1log_2(JN_1)) $.
{\color{black}The following calculation of $ \hat{\bf{x}}_1^l $ contains an inverse matrix, i.e., $ ( {{{1}/{\sigma_1^2}{\bf I}_{K_1} + {\eta_1^2\rho{\bf F}_1^H{\bf F}_1 }}})^{-1} $.
However, we can see that it can be calculated offline in advance.}
Similarly, the calculation of $ {\bf{v}}_2^l $ in $ \hat{\bf{x}}_2 $-minimization step roughly requires twice $ JN_2 $-point FFT operation, which leads to $ \mathcal{O}(2JN_2log_2(JN_2)) $ computation complexity.
Thus, the computational cost to obtain $ \hat{\bf{x}}_1^{l+1} $ and $ \hat{\bf{x}}_2^{l+1} $ is dominant by $ JN_1 $ and $ JN_2 $-point FFT operations, i.e. $ \mathcal{O}(JN_1log_2(JN_1)+2JN_2log_2(JN_2)) $.
In the $ \hat{\bf{z}} $-minimization step, the calculation of $ {\bf{u}}^l $ equally requires once $ JN_1 $-IFFT and twice $ JN_2 $-IFFT operations.
The incurred complexities are $ \mathcal{O}(JN_1log_2(JN_1))$ and $ \mathcal{O}(2JN_2log_2(JN_2)) $, respectively.
From (\ref{clip_u}), we can also see that every entry $ \hat{z}_i^{l+1} $ in $ \hat{\bf{z}}^{l+1} $ can be easily computed through once multiplication.
In addition, since $ {{\bf{F}}_1}\hat{\bf{x}}_1^{l + 1} + {{\bf{F}}_2}\hat{\bf{x}}_2^{l + 1} $ has been calculated in the $ \hat{\bf{z}} $-minimization step, the dual variable $ \hat{\bf{y}}^{l + 1} $ can be obtained merely through $ L\!_{s\!y\!s} $ complex multiplications.
According to the analysis above, we can conclude that the total computational cost in each iteration of O-ADMM is about $ \mathcal{O}(JN_1log_2(JN_1)+2JN_2log_2(JN_2)) $.
On the other hand, if the SOCP model in (\ref{originOp}) is solved by standard interior-point methods, the computation complexity of solving the optimization problem is $ \mathcal{O}((N_1+2N_2)^3) $\cite{2011ITC-WangOptimized}.
Because of $ N\gg J $ in most cases, 
the proposed O-ADMM algorithm is highly efficient to solve the original optimization problem (\ref{originOp}) compared with traditional interior-point methods.

As for the CU-ADMM algorithm, the only difference from O-ADMM is that the constraint in the $ \hat{\bf{z}} $-minimization step is updated in each iteration.
This extra computational cost is $ O(L\!_{s\!y\!s}) $, which is far less than the FFT implementation.
Hence, the computation complexity of CU-ADMM is also $ \mathcal{O}(JN_1log_2(JN_1)+2JN_2log_2(JN_2)) $.

For comparison, we also consider the complexity of the NS-ICF method.
The calculation of the clipping level incurs a complexity of $ O(L\!_{s\!y\!s}) $.
Besides, The clipping procedure also leads to $ O(L\!_{s\!y\!s}) $.
One can also find that the filtering procedure results in the primary computation complexity, which contains once $ JN_1 $-point and twice $ JN_2 $-point FFT/IFFT pair operations.
Consequently, the computation complexity of NS-ICF is roughly $ \mathcal{O}(JN_1log_2(JN_1)+2JN_2log_2(JN_2)) $.
{\color{black}
The complexity comparison of the interior-point method, the proposed ADMM-based algorithms and the NS-ICF method is summarized in Table \ref{table2}.
Based on the analysis above, we can see that using interior-point method to solve SOCP has the highest computational cost.
On the other hand, the complexities of the O-ADMM and CU-ADMM algorithms are much lower than interior-point method, in which each iteration has comparable computation complexity to the NS-ICF method.
}

\begin{table}[!t]
	\renewcommand{\arraystretch}{1.1}
	\caption{{\color{black}Complexity comparison of the interior-point method, ADMM-based algorithms and NS-ICF method}} \label{table2} \centering
	\begin{tabular}{c|c}
		\hline
		{Algorithm}&{Computational Complexity}\\
		\hline
		{interior-point method}&$ \mathcal{O}((N_1+2N_2)^3) $\\
		\hline
		{O-ADMM/CU-ADMM}&$ \mathcal{O}(JN_1log_2(JN_1)+2JN_2log_2(JN_2)) $\\
		\hline
		{NS-ICF}&$ \mathcal{O}(JN_1log_2(JN_1)+2JN_2log_2(JN_2)) $\\
		\hline
	\end{tabular}
\end{table}

\begin{figure*}[t]
	\centering
	\includegraphics[width=120mm]{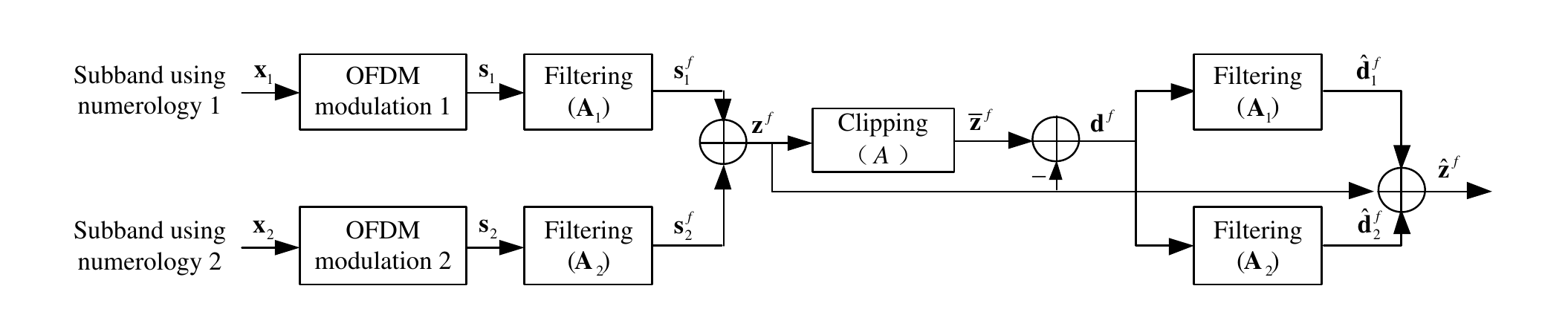}
	\caption{{\color{black}Illustration of the NS-ICF method with filtering.}}
	\label{clip3}
\end{figure*}

\subsection{Convergence Issue and Algorithm Summary}
The ADMM approach has been widely used to solve structured convex optimization problems.
However, most of the applications of ADMM is limited to the case of two separable variables, such as \cite{2018ITVT-BaoADMM,2018ITSP-YongchaoWangOptimized}, in which the algorithm has the theoretical linear convergence rate.
The convergence of the ADMM with more than two variables remains an open question for a long time.
In our cases, the mixed-numerology system always deal with $ M\geqslant2 $ subbands, which  leads to more than two variables in the optimization models (\ref{Op2}).
Fortunately, the recent work \cite{lin2015global} shows that under three easily verifiable conditions the convergence of the ADMM with more than two variables still holds.
As we can see from the (\ref{Op2}), the objective function involves $ M $ 2-norm functions that can be easily proven to be strongly convex and Lipschitz continuous.
Meanwhile, the linear constraint (\ref{Op2,b}) shows that the coefficient matrices  of $ \hat{\bf{x}}_1 $, $ \hat{\bf{x}}_2 $ and $ \hat{\bf{z}} $ have full column rank.
The conditions in \cite{lin2015global} are then satisfied and our proposed O-ADMM is hence convergent as well.
Interested readers are referred to \cite{lin2015global} for the detailed discussion.
As for the CU-ADMM, the step of the constraint update is heuristic and the theoretical proof for this modification is not straightforward.
However, the simulation results show that the convergence can be always ensured.

According to the iterative steps of the proposed algorithms, we can see that the subproblems of $ \hat{\bf{x}}_1 $ and $ \hat{\bf{x}}_2 $ are performed on the subbands signals and the subproblem of $ \hat{\bf{z}} $ produces the optimized composite signal.
From this perspective, the ADMM framework can be seen as the generating procedure of PAPR reduced composite signal.
Therefore, the proposed ADMM-based algorithms can be extended to the cases that have any number of numerologies.

{\color{black}
\section{Applications with Filtering and Windowing Techniques}
In order to improve the spectrum utilization efficiency as well as suppress the OOBE, filtering and windowing techniques are proposed as two main waveform enhancements in 5G NR\cite{2017ICM-Lien5G}.
Filtering operation is performed on each subband with predesigned frequency response, which is also known as F-OFDM\cite{2016ICM-ZhangWaveform}.
Windowed-OFDM (W-OFDM) introduces a well-chosen window function to be multiplied with the boundaries of each OFDM symbol\cite{2016ICM-Waveform}.
F-OFDM and W-OFDM inherit the benefits of traditional OFDM while achieving the spectrum confinement, both of which are simple and can be transparent to the receiver.
However, F-OFDM and W-OFDM maintain the high PAPR problem and thus suffer from PA nonlinearity.
Nonlinear distortion also results in spectral regrowth that degrades the expected spectrum confinement performance\cite{2017ICM-Lien5G}.

In this section, we will show that the basic idea behind the proposed PAPR reduction methods can be easily extended to the variants of OFDM.
In combination with filtering and windowing techniques,
the proposed PAPR reduction methods are able to preserve the suppression of OOBE while reducing the PAPR.

\subsection{NS-ICF with Filtering}

We assume that the $ i $-th subband-wise filter impulse response is
$ {\bf a}_i = [a_i(0),a_i(1),...,a_i(L_f-1)] $, 
where $ L_f $ is the filter length.
In addition, the prototype filter design used in each subband is assumed to be same. 
We let the energy of $ {\bf a}_i $ be normalized to unity, i.e., $ \sum_{l=0}^{L_f-1}|a_i(l)|^2 =1$.
In order for matrix representation, we use an $ (L_{s\!y\!s}+L_f-1) \times L\!_{s\!y\!s} $ dimension Toeplitz matrix $ {\bf A}_i $ consisting of the filter impulse response to realize linear convolution.
Then, the filtered signal in the $ i $-th subband can be expressed as 
\begin{equation}\label{}
{{\bf{s}}_i^{f}} = {\bf A}_i{{\bf{F}}_i}{{\bf{x}}_i}.
\end{equation}
Similarly, for $ M=2 $, the composite signal $ {{\bf{z}}_i^{f}} $ composed of filtered subbands signals can be written as
\begin{equation}\label{}
{{\bf{z}}^{f}} = {\bf A}_1{{\bf{F}}_1}{{\bf{x}}_1} + {\bf A}_2{{\bf{F}}_2}{{\bf{x}}_2}.
\end{equation}

The idea behind the NS-ICF method is to use the clipping noise instead of clipped signal to the filter, which avoids INI accumulation during the repeated execution.
Combined with F-OFDM, the filtering procedure in NS-ICF can be realized through subband-wise filter.
Fig. \ref{clip3} illustrates the processing structure of NS-ICF for F-OFDM.
It can be seen that after the clipping procedure, clipping noise is applied to individual subbands filters.
The filtered clipping noise is obtained as $ {{\bf{\hat d}}_i^{f}} = {{\bf{A}}_i^{f}}{{\bf{d}}^{f}} $ for $ i = 1,2 $.
Thus, the output composite signal is
\begin{equation}\label{z_clip3}
{{\bf{\hat z}}^{f}} = {{\bf{ z}}^{f}} + \left( {{\bf{ F}}_1}{{\bf{ A}}_1} + {{\bf{ F}}_2}{{\bf{ A}}_2}\right){{\bf{d}}^{f}}.
\end{equation}
Similar to (\ref{z_clip2}), we observe that the INI accumulation is also avoided in (\ref{z_clip3}).

Compared with the original NS-ICF method, the filtering procedure employs the time-domain filter of F-OFDM. 
Thus, the computational cost mainly depends on the time domain convolution that leads to $ \mathcal{O}(L_{\!f} L_{\!s\!y\!s}) $ complexity.
For F-OFDM, since the filter length can be set as half of the signal
duration to achieve fast frequency roll-off,
the computation complexity of F-OFDM is significantly larger than original OFDM.
Hence, we find that although the NS-ICF method combined with filtering technique provides better spectrum confinement, the computational cost is greatly increased.

\subsection{Optimization Model with Windowing}
The transmitter windowing $ w\left( n \right) $ is commonly generated by the raised cosine shape function in \cite{2016ICM-Waveform}.
The windowing function has unit response in the middle and smoothly converges to zero on the two sides.
The roll-off factor is $ \beta \in [0,1) $ that controls the length of roll-off portion of the window, i.e., $ L_{\!r\!o\!f\!f} = \beta L_{sys} $.
$ L_{\!r\!o\!f\!f} $ samples at the beginning of the CP are weighted by the increasing window while an extra cyclic postfix with $ L_{\!r\!o\!f\!f} $ samples added after the OFDM symbol is multiplied by the decreasing window.

When transmitter windowing is applied, subband signal can be expressed as 
\begin{equation}\label{}
{{\bf{s}}_i^{w}} = {\bf W}_i{{\bf{P}}_i^w}{{\bf{D}}_i}{{\bf{x}}_i},
\end{equation}
where the diagonal matrix $ {\bf W}_i=diag(w_i(0),...,w_i(L_{\!s\!y\!s}+L_{\!r\!o\!f\!f}-1)) $ denotes the windowing operation and $ {\bf{P}}_i^w =  [{\bf{P}}_i;{\bf{I}}_{L_{\!r\!o\!f\!f}},{\bf{0}}_{L_{\!r\!o\!f\!f}\times ({L_{\!s\!y\!s}-L_{\!r\!o\!f\!f}})}] $ performs the cyclic extension at both the prefix and postfix.
Then, by denoting $ {{\bf{F}}_i^w} = blkdiag(\eta_i{\bf W}_i{{\bf{P}}_i^w}{{\bf{D}}_i},2^{v_i}) $, the composite signal for $ M=2 $ can be written as
\begin{equation}\label{z3}
{{\bf{z}}^{w}} = {{\bf{F}}_1^w}{{\bf{x}}_1} + {{\bf{F}}_2^w}{{\bf{x}}_2}.
\end{equation}
We can find that the only difference between (\ref{z3}) and (\ref{vector z}) is the coefficient matrix $ {{\bf{F}}_i^w} $.
Thus, this expression can be easily used in the optimization model (\ref{Op2}) to take the place of the constraint (\ref{Op2,b}),
which can be formulated as
\begin{subequations}
	\label{Op3}
	\begin{align}\label{Op3,a}
	\mathop {{\rm{min}}}\limits_{{{\hat {\bf{x}}}_1},{{\hat {\bf{x}}}_{2}},\hat{\bf{z}}^w} \quad
	&\frac{1}{2\sigma_1^2}{{\left\| {{{\bf{x}}_1} - {{{\bf{\hat x}}}_1}} \right\|_2^2}} + \frac{1}{2\sigma_2^2}{{\left\| {{{\bf{x}}_2} - {{{\bf{\hat x}}}_2}} \right\|_2^2}}\\
	\label{Op3,b}
	{s.t.} \quad
	&\hat{\bf{z}}^w = {{{\bf{F}}_1^w}{\hat {\bf{x}}}_1 + {\bf{F}}_{2}^w{\hat {\bf{x}}}_{2}},
	\\
	\label{Op3,c}
	& {{\left\| \hat{\bf{z}}^w \right\|_\infty }} \le \gamma \frac{1}{\sqrt{L_{\!s\!y\!s}}}{\left\| {{\bf{z}}^w} \right\|_2}.
	\end{align}
\end{subequations}

Note that the constraint (\ref{Op3,b}) still satisfies a linear relationship among $ {\hat {\bf{x}}}_{1} $, $ {\hat {\bf{x}}}_{2} $ and $ \hat{\bf{z}}^w $.
Therefore, the proposed O-ADMM and CU-ADMM algorithms can also be applied to solve this optimization problem (\ref{Op3}). 
In addition, the windowing operation only needs to multiply the samples residing on the edges of signal by the increasing and decreasing window function, so that the windowing technique generally has negligible complexity overhead compared with the filtering technique.
Hence, the computation complexities of ADMM-based algorithms combined with windowing technique are similar to those without windowing.
}

\begin{figure}[t]
	\centering
	\includegraphics[width=70mm]{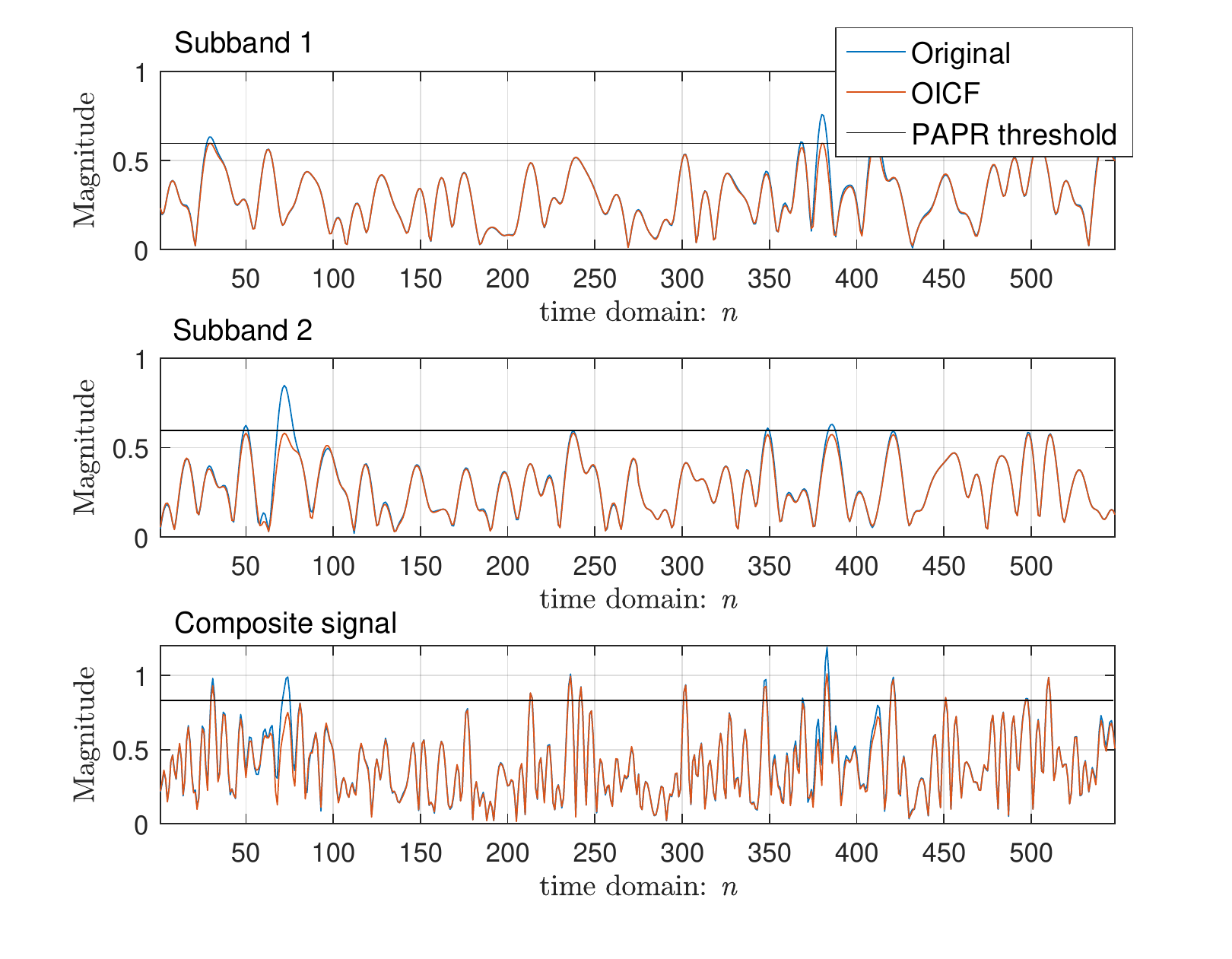}
	\caption{Signal magnitude when using OICF method on individual subbands.}
	\label{figure3}
\end{figure}
\begin{figure}[t]
	\centering
	\includegraphics[width=70mm]{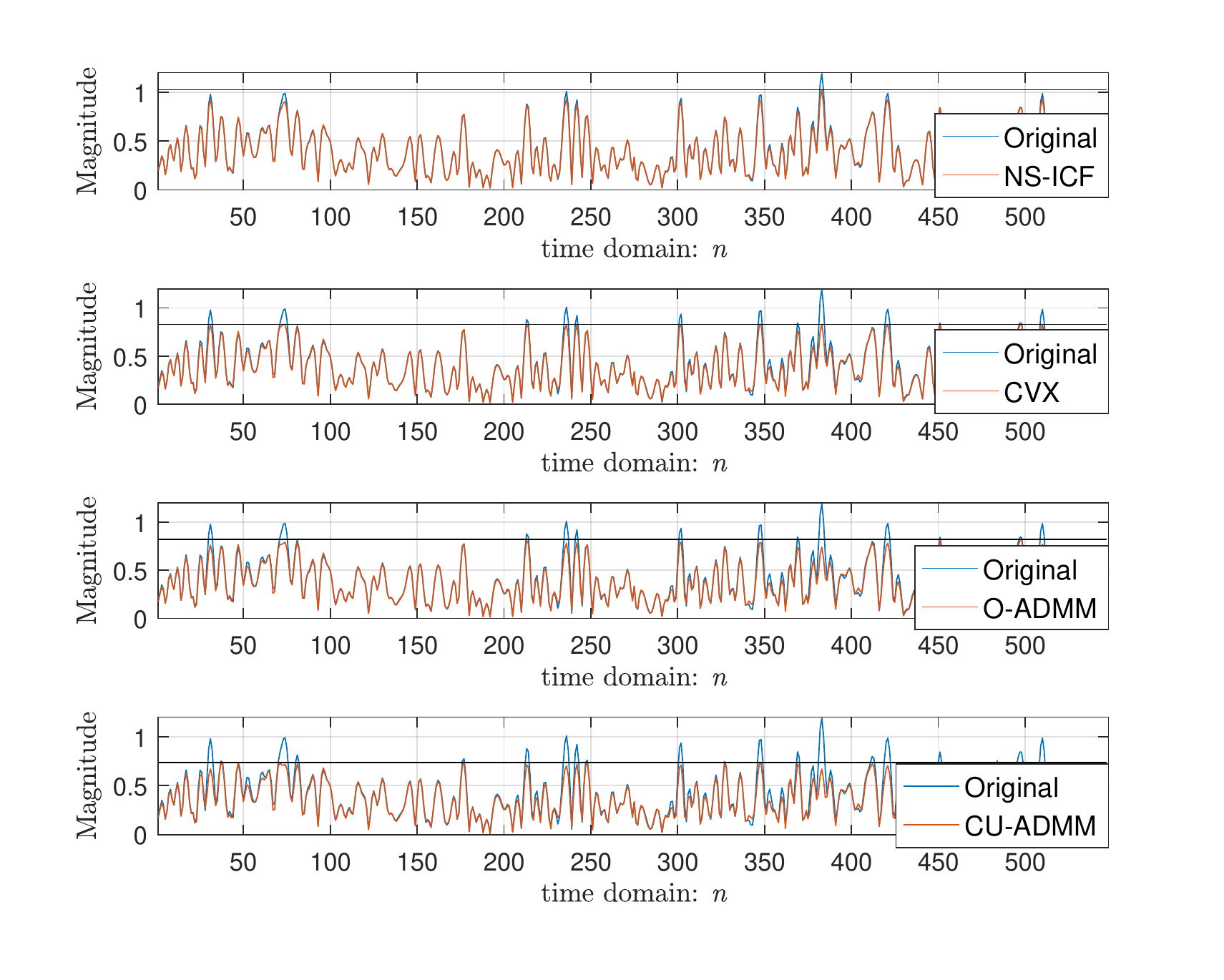}
	\caption{Magnitude of original and PAPR-reduced composite signals using the proposed NS-ICF method, CVX, O-ADMM and CU-ADMM algorithm.}
	\label{figure4}
\end{figure}

\section{Simulation Results}
In this section, we present simulation results to evaluate the performance of the proposed NS-ICF method and optimization method (the O-ADMM and CU-ADMM algorithms).
In the simulations, we consider two adjacent subbands with different selection of numerology as illustrated in Fig. \ref{fig2}, i.e., $ f_1 = {1}/{2}f_2 $.
{\color{black}The CP length takes 7\% of the symbol period.}
The first subband modulates the data symbols on 56 subcarriers, while the second subband has $ 28 $ data subcarriers.
Meanwhile, the reserved guard band is $ G_1=8f_1 $ between the subband 1 and subband 2.
The oversampling rate is set as $ J=4 $ and the QPSK modulation is adopted by both of subbands.
The penalty parameter for the proposed ADMM-based algorithms is $ \rho = 0.25 $.
The CR is set to 5dB.
The transmission power on every subband is equal to each other and normalized to 1, which makes the total power evenly allocated on the available bandwidth.
In our simulation, we randomly generate 5000 LCM symbols.



\subsection{PAPR Reduction Verification}

Firstly, we examine the time-domain composite signal when using the OICF method on separate subbands.
The original signal is also depicted for comparison.
As shown in Fig. \ref{figure3}, the first and second plots are the signal amplitudes in subband 1 and 2, respectively.
It can be observed that the peaks can be reduced compared with the original signal in each subband.
Nonetheless, when combining two subbands' signals, 
several peaks still rise high in the third sub-figure of Fig. \ref{figure3}.
This demonstrates that the PAPR reduction techniques individually applied to each subband do not work in the mixed-numerology systems.
In Fig. \ref{figure4}, we show the signal amplitudes generated by the NS-ICF method and optimization method.
The optimization problem for the SOCP in (\ref{originOp}) is solved by the proposed O-ADMM algorithm and CVX individually.
We can see that all methods are able to reduce the PAPR of composite signal.
Specifically, {\color{black}the horizontal lines indicate that} the NS-ICF method achieves PAPR of 7dB, while the optimization methods using CVX and O-ADMM attain exactly the same PAPR of 5.5dB.
Furthermore, it is of interest that the proposed CU-ADMM algorithm achieves the best PAPR performance, which reaches to the desired PAPR threshold 5dB within one execution.

\subsection{Algorithm Efficiency of O-ADMM and CU-ADMM}
\begin{figure}[t]
	\centering
	\includegraphics[width=68mm]{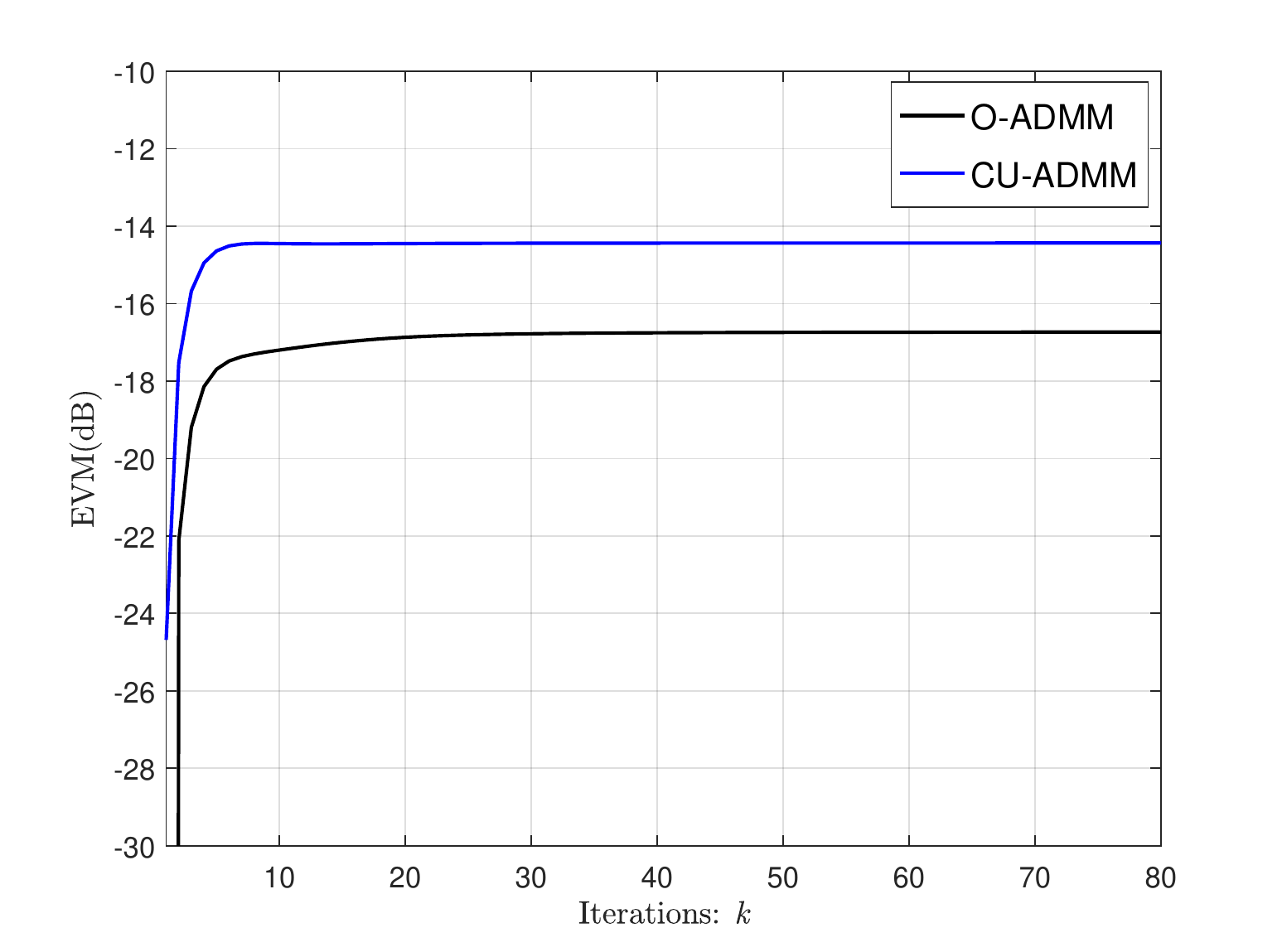}
	\caption{Objective function with respect to the number of iterations.}
	\label{obj}
\end{figure}
\begin{figure}[t]
	\centering
	\includegraphics[width=68mm]{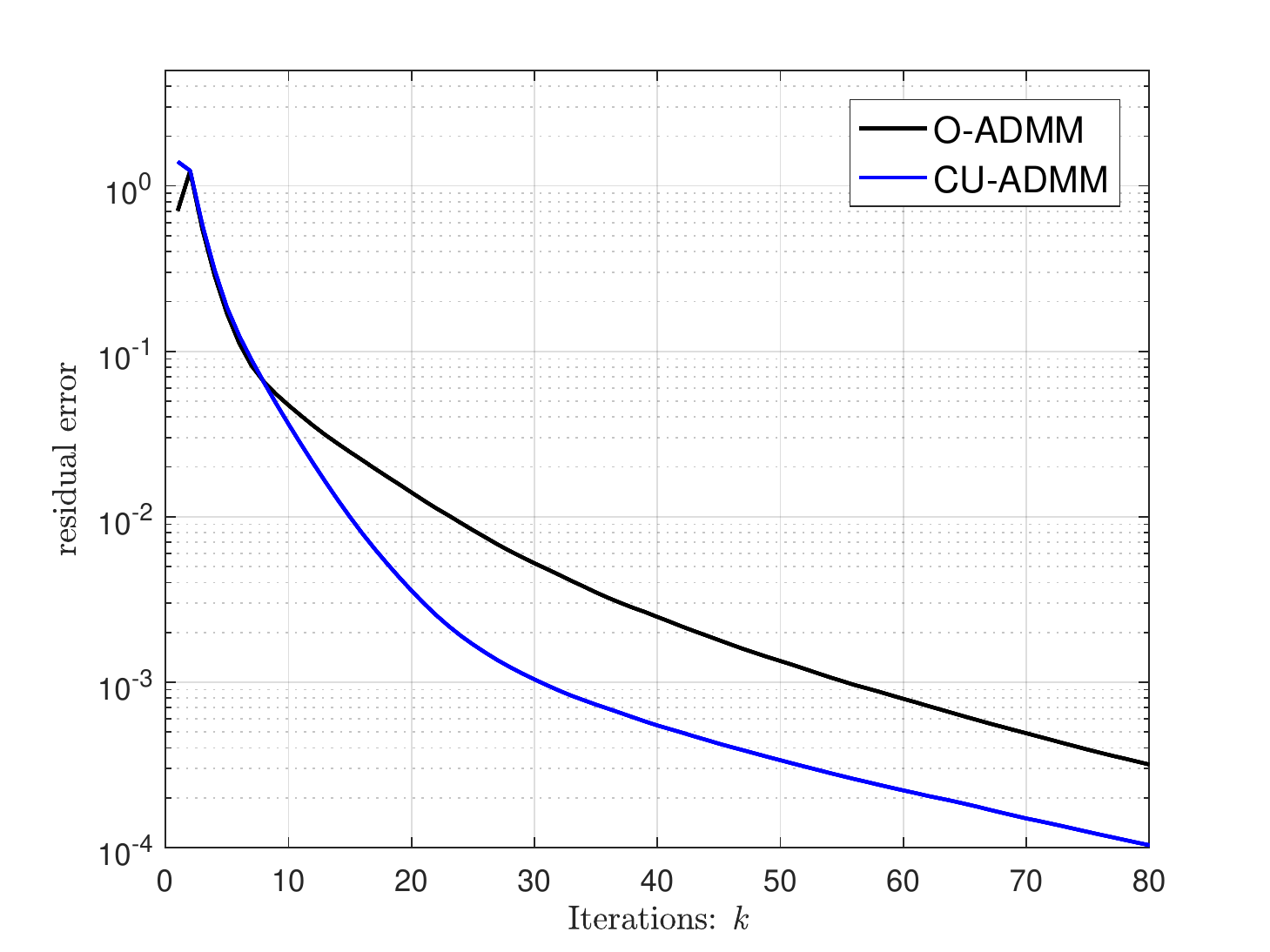}
	\caption{Primal residual with respect to the number of iterations.}
	\label{resdual}
\end{figure}
To evaluate the convergence of the proposed O-ADMM and CU-ADMM algorithms,
we plot the EVM of composite signal with respect to the number of iterations.
In Fig. \ref{obj}, we can see that both O-ADMM and CU-ADMM algorithms can converge with no more than 10 iterations.
We can also find that the resulting EVM of CU-ADMM is slightly worse than O-ADMM due to the iterative update of the constraint in the $ \hat{\bf z} $-minimization step.
In Fig. \ref{resdual}, the primal residual error is defined as $ \| {{{\bf{F}}_1}\hat{\bf{x}}^l_1 + {{\bf{F}}_{2}}\hat{\bf{x}}_{2}^l - {\hat{\bf{z}}^l}} \|_2^2 $.
From the convergence curves, one can observe that the residual error decreases rapidly in the first few times and this bend becomes gradually flat after 10 iterations.
Meanwhile, the CU-ADMM algorithm shows better convergence rate than O-ADMM.
According to the two figures, the stopping criterion can be determined carefully. 

Apart from the convergence, the low computation complexity of the proposed ADMM-based algorithms is also one major reason that leads to high efficiency.
Table \ref{table} compares the running time of the ADMM-based algorithm with 10 iterations to that of using the CVX toolbox.
The simulation environment is based on MATLAB.
According to the execution times shown in Table \ref{table}, the proposed ADMM-based algorithms are nearly 1000 times time-saving compared with the SOCP model using CVX, which is of significant benefits for practical implementation.
In addition, the elapsed time of CU-ADMM is slightly higher than that of O-ADMM, which is probably due to the constraint update step.

\begin{table}[!t]
	\renewcommand{\arraystretch}{0.9}
	\caption{Execution time comparison of O-ADMM, CU-ADMM, and CVX methods} \label{table} \centering
	\begin{tabular}{c|c|c|c}
		\hline
		{Algorithm}&{O-ADMM}&{CU-ADMM}&{CVX}\\
		\hline
		{Time elapse (ms)}&6.1 &6.2&6249\\
		\hline
	\end{tabular}
\end{table}

\subsection{PAPR Comparison}

\begin{figure}[t]
	\centering
	\includegraphics[width=70mm]{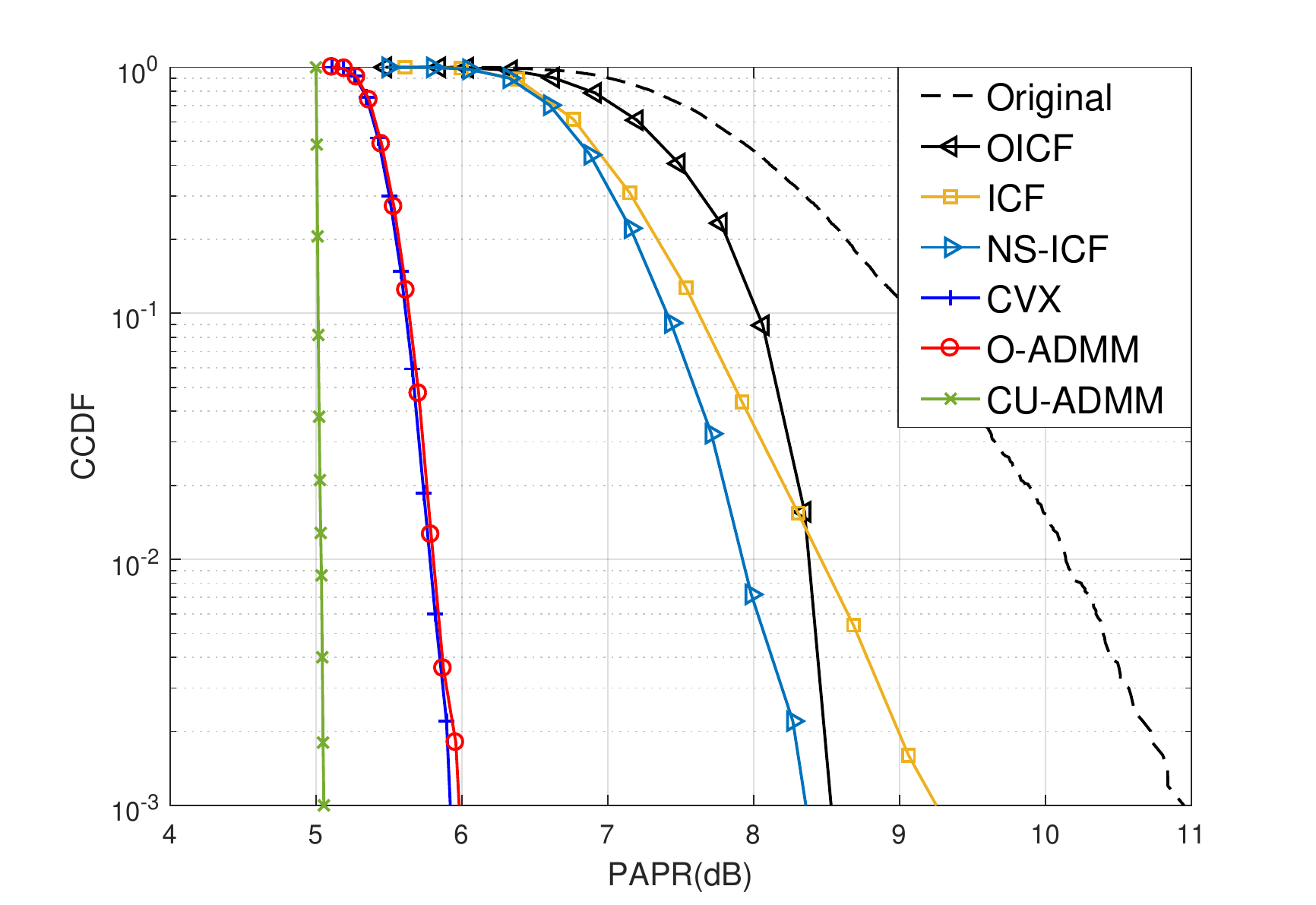}
	\caption{PAPR performance of the ICF, NS-ICF methods, O-ADMM and CU-ADMM algorithms, CVX, and OICF method when the CR is set to 5dB.}
	\label{CCDF1}
\end{figure}
\begin{figure}[t]
	\centering
	\includegraphics[width=70mm]{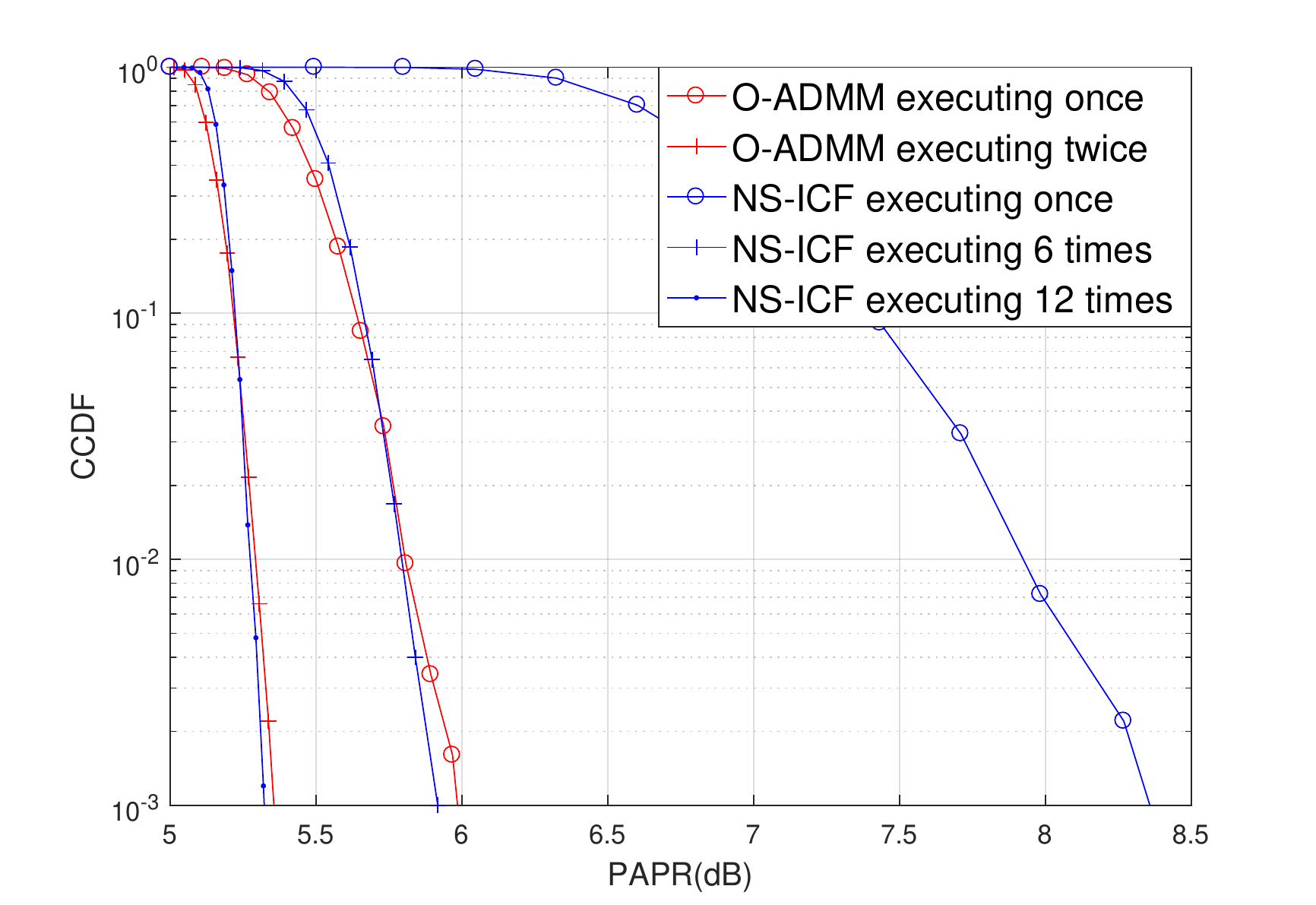}
	\caption{{\color{black}PAPR performance of the NS-ICF and optimization methods with different times of execution.}}
	\label{CCDF2}
\end{figure}

\begin{figure}[t]
	\centering
	\includegraphics[width=70mm]{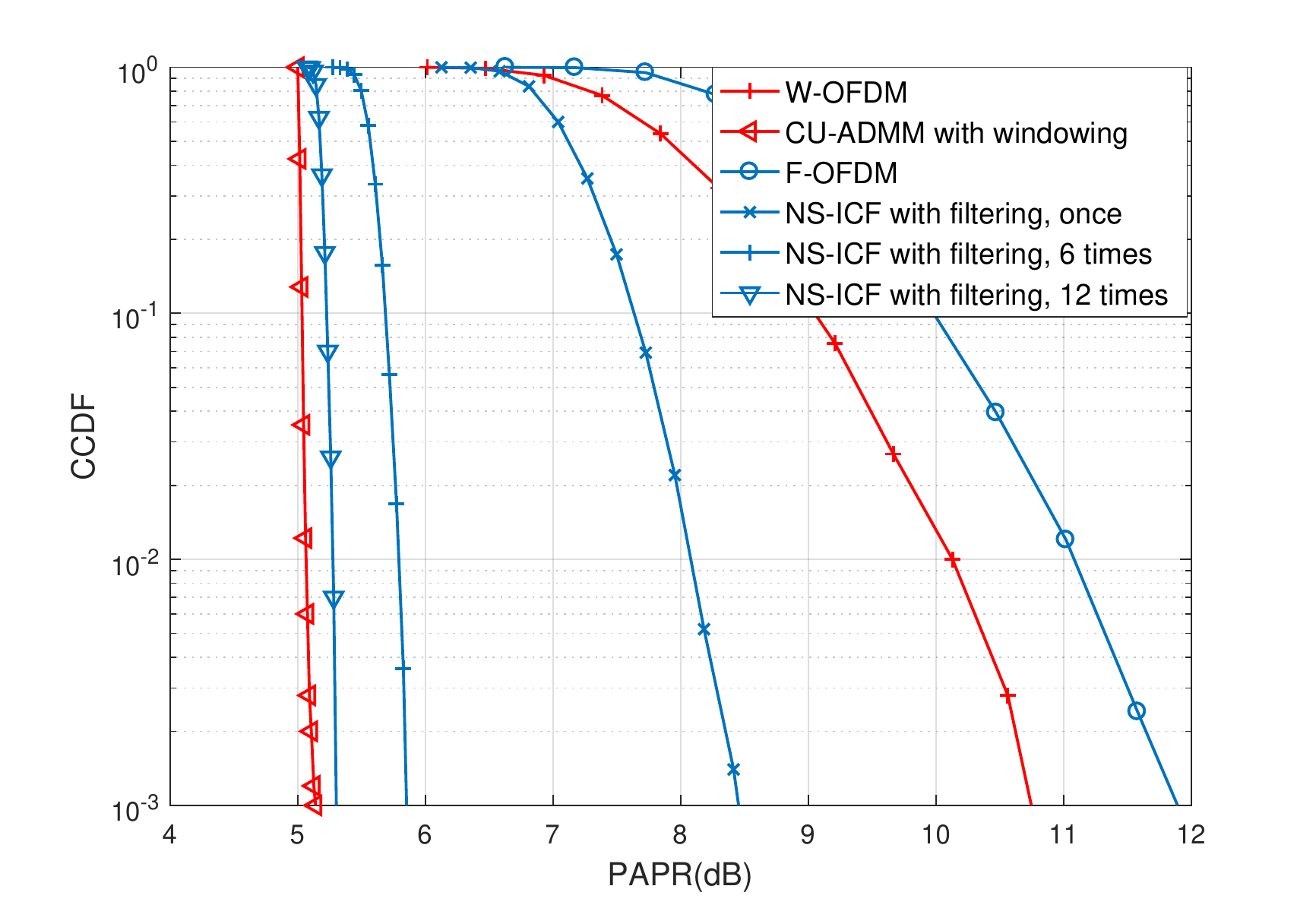}
	\caption{{\color{black}PAPR performance of the NS-ICF method with filtering and CU-ADMM algorithm with windowing.}}
	\label{CCDF3}
\end{figure}
\begin{figure}[t]
	\centering
	\includegraphics[width=80mm]{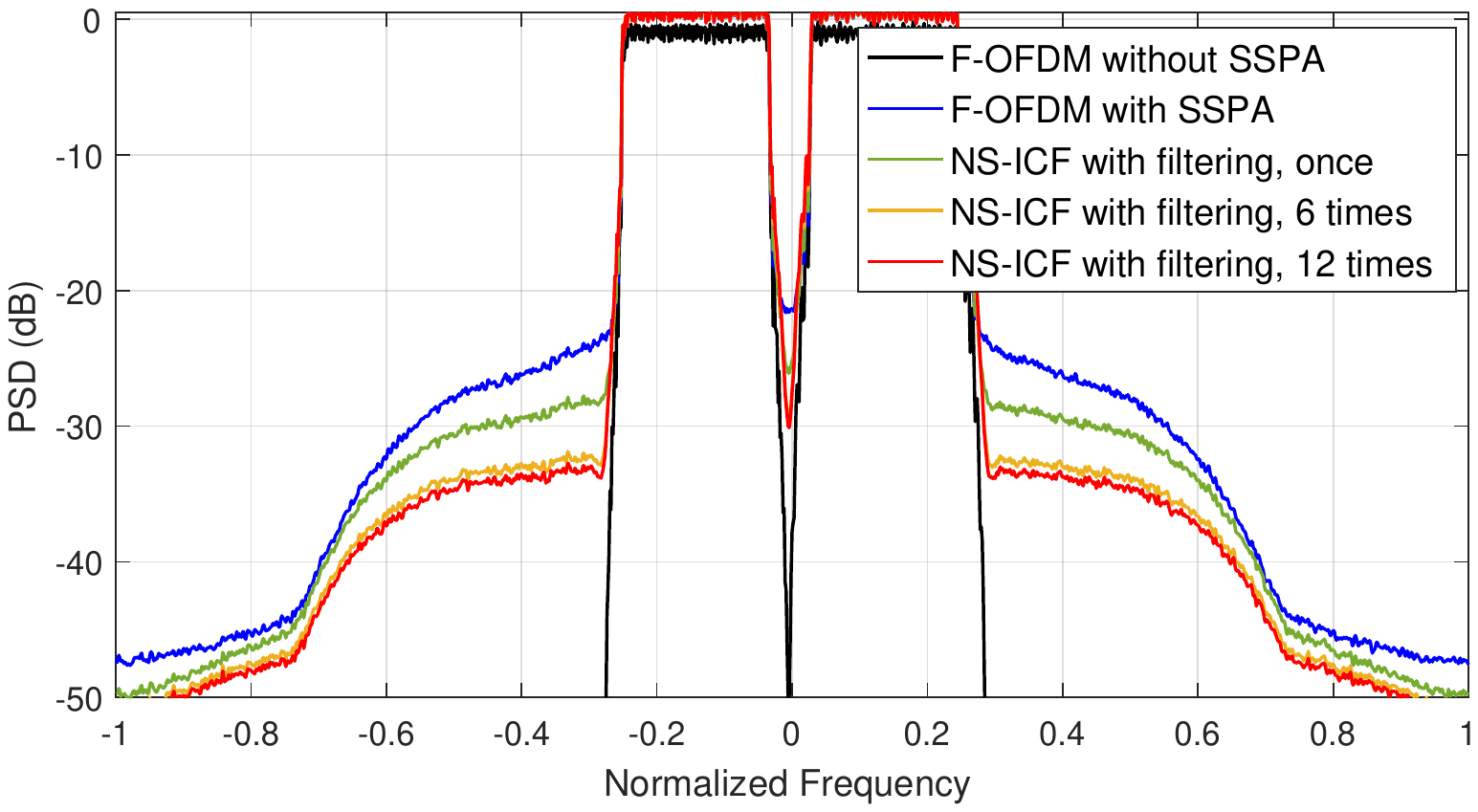}
	\caption{{\color{black}OOBE performance of the NS-ICF method with filtering.}}
	\label{PSD1}
\end{figure}
\begin{figure}[t]
	\centering
	\includegraphics[width=80mm]{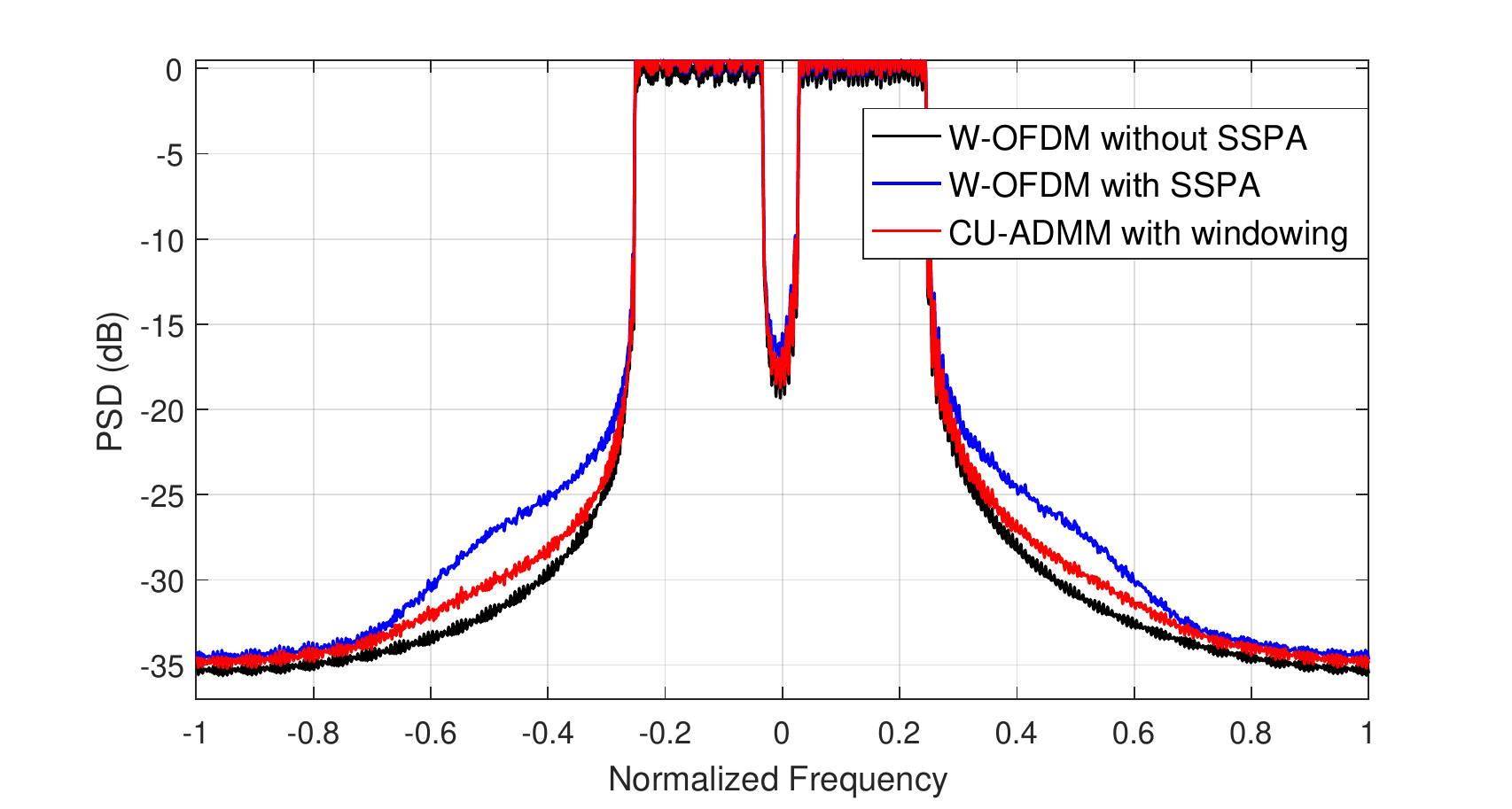}
	\caption{{\color{black}OOBE performance of the CU-ADMM algorithm with windowing.}}
	\label{PSD2}
\end{figure}

In this subsection, we use the complementary cumulative distribution function (CCDF) to evaluate the PAPR reduction performance.
It is defined as the probability that the PAPR of signal exceeds a given PAPR value, which is expressed as
\begin{align} 
\rm{CCDF} = \rm{Pr}(\rm{PAPR}>\rm{PAPR}_0),
\end{align}
where $ \rm{Pr}(\cdot) $ denotes the probability function.

Fig. \ref{CCDF1} shows the CCDF curves of different PAPR reduction methods with one execution. 
The CCDF of the original composite signal is also considered as a reference.
As expected, the OICF method working on the separate subbands has the worst PAPR performance, which only has 2dB PAPR reduction compared to the original signal.
{\color{black}Besides, the direct application of the ICF method leads to a slow fall of CCDF curve.}
Meanwhile, it can be seen that the proposed NS-ICF method leads to limited improvement of PAPR performance.
Due to the peak regrowth, the NS-ICF method, like the classical ICF method, usually has to be executed many times to gradually approach the desired PAPR.
The optimization method, however, achieves significant performance of PAPR reduction and shows steep curves of CCDF.
To be more specific, when solving the optimization problem (\ref{Op}), the proposed O-ADMM algorithm and the CVX toolbox have visually overlapped CCDF curves, which are very close to the preset PAPR threshold.
However, due to the high efficiency of the ADMM approach, the O-ADMM algorithm is far more computationally efficient. 
Furthermore, we can observe that the proposed CU-ADMM algorithm shows a cut-off CCDF curve right on the desired PAPR.
It shows that updating the constraint of $ \hat{\bf{z}} $-minimization step at each iteration leads to the further improvement on PAPR performance.

Fig. \ref{CCDF2} plots the CCDFs of the PAPR using NS-ICF method (executing 1, 6, and 12 times), and the proposed O-ADMM algorithm (executing 1 and 2 times).
It can be seen that both the NS-ICF method and the O-ADMM algorithm can gradually reduce the PAPR to the desired level with being executed repeatedly.
However, for the O-ADMM algorithm, the PAPR reduction is about 5.9dB at a probability of $ 10^{-3} $ after executing once and about 5.3dB after executing twice,
while the NS-ICF method requires 6 and 12 times of execution to achieve a comparable PAPR reduction performance.
Thus we can reach a similar conclusion as in \cite{2011ITC-WangOptimized} that the optimization method requires fewer executions to approach the PAPR threshold.

{\color{black}

To illustrate the applications of proposed PAPR reduction methods with spectrum confinement techniques, F-OFDM and W-OFDM are considered in our simulations.
Subband-wise filters are designed based on the root raised cosine windowed Sinc function with 128 length for F-OFDM.
The raised cosine window with $ \beta=0.04 $ is applied for W-OFDM.
Meanwhile, we use the NS-ICF method for F-OFDM and the CU-ADMM algorithm for W-OFDM.
From Fig. \ref{CCDF3}, it can be observed that the proposed methods are still effective to reduce the PAPR of composite signal.
The NS-ICF method results in more than 6dB PAPR reduction after 12 times executions.
Besides, as expected, the CU-ADMM algorithm achieves a steep drop at the PAPR of 5dB.

In order to show the influence of PAPR reduction methods on the OOBE of F-OFDM and W-OFDM, we further consider passing the composite signal through the solid state power amplifier (SSPA)\cite{2013ICST-RahmatallahPeak}.
The smoothing factor of SSPA model is set to 3 and the input back-off (IBO) is 5dB.
The power spectrum densities (PSD) are depicted in Fig. \ref{PSD1} and Fig. \ref{PSD2}, which are computed by means of periodogram.
From Fig. \ref{PSD1}, we can see that the OOBE performance of F-OFDM is obviously degraded at the output of SSPA.
When using the proposed NS-ICF method, it can achieve near 10dB lower OOBE than that without PAPR reduction after 12 times of execution.
As for W-OFDM, the proposed CU-ADMM algorithm leads to a similar OOBE performance compared to the W-OFDM without SSPA in Fig. \ref{PSD2}.
Note that spectrum confinement is of great importance in 5G communications.
Thus, it is beneficial to exploit the PAPR reduction techniques to avoid the nonlinearity of PA, which definitely saves the property of low OOBE brought by spectrum confinement techniques.
}

\subsection{EVM Analysis}
We now show the comparison of distortion for different PAPR reduction methods.
The root mean-square (RMS) EVM is used here to evaluate the distortion statistically, which is defined as 
\begin{equation}\label{}
{\rm{RMS\,EVM}}=\sqrt{E[{\rm{EVM}^2}]},
\end{equation}
where the expectation is calculated by averaging 5000 LCM symbols.
The target PAPR threshold is 5dB for the CU-ADMM algorithm and it is carefully chosen for ICF, NS-ICF and O-ADMM so that the same PAPR can be achieved with one execution.

\begin{table}[!t]
	\renewcommand{\arraystretch}{0.9}
	\caption{Performance comparison of the ICF, NS-ICF methods and optimization method (using O-ADMM, CU-ADMM algorithms).} \label{EVM2} \centering
	\begin{tabular}{c|c|c|c}
		\hline
{Algorithm}&{\makecell{EVM(dB) in \\Subband 1}}&{\makecell{EVM(dB) in\\ Subband 2}}&{\makecell{EVM(dB) of \\LCM symbol}}\\
\hline
{\color{black}{ICF}}&-13.75&-13.74&-10.73\\
\hline
{NS-ICF}&-15.50&-15.51&-12.50\\
\hline
{O-ADMM}&-17.25&-17.25&-14.24\\
\hline
{CU-ADMM}&-17.04&-17.04&-14.03\\
\hline
	\end{tabular}\label{EVM_TABLE}
\end{table}
From Table \ref{EVM_TABLE}, we can see that the optimization method achieves better EVM performance than the ICF and NS-ICF methods.
The O-ADMM algorithm achieves the EVM of -14.24dB, while the CU-ADMM only results in negligible increase of EVM, which is -14.03dB.
The NS-ICF method generally causes EVM deterioration, which is 1.74dB worse than the O-ADMM algorithm.
{\color{black}
Moreover, the EVM performance of the ICF method is further degraded due to the INI accumulation.}
Additionally, we can see that the distortion is fairly uniform on both subbands, each of which has similar EVM performance to the other.

\begin{figure}[t]
	\centering
	\includegraphics[width=70mm]{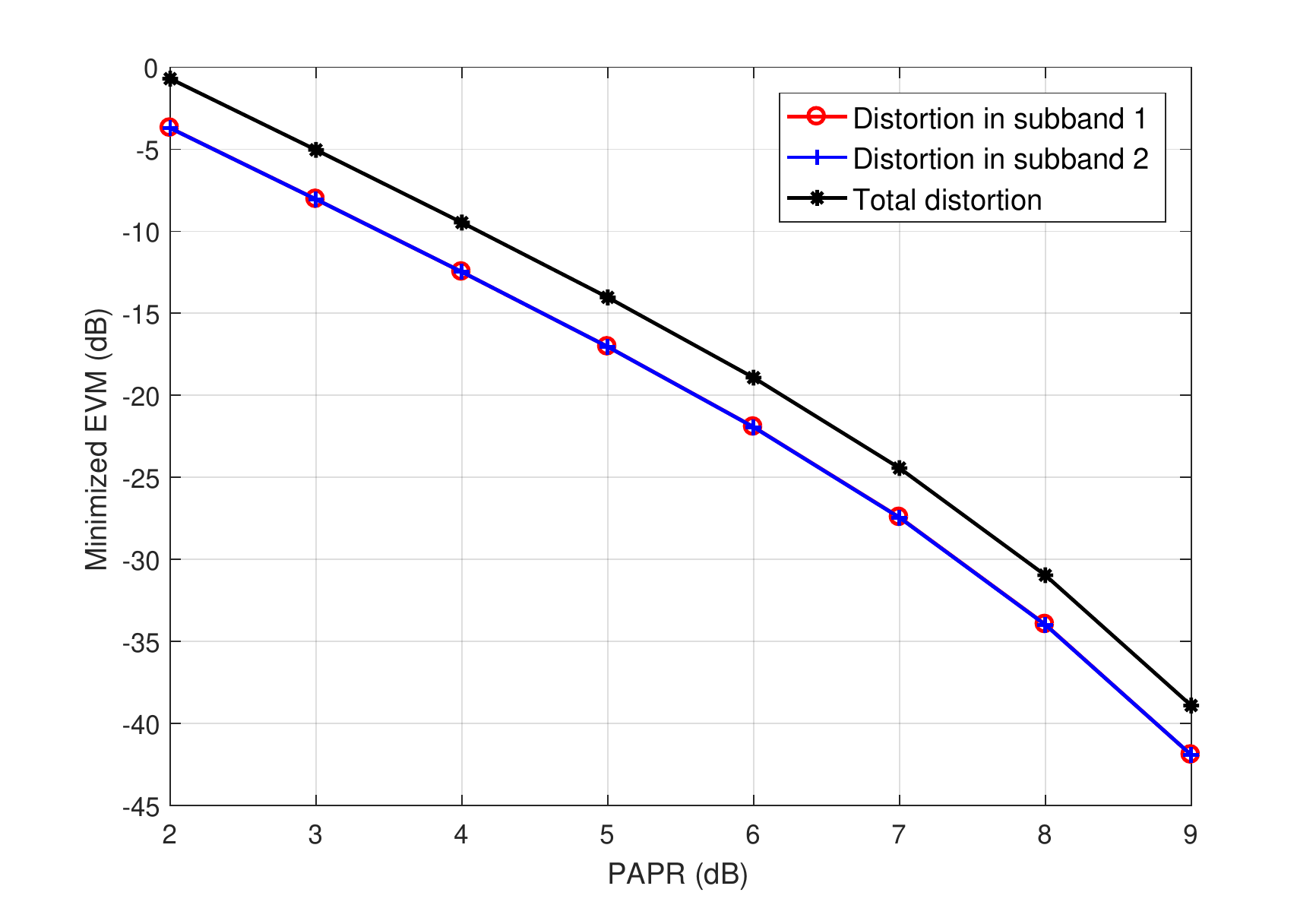}
	\caption{Relation between the target PAPR and EVM performance using the CU-ADMM algorithm.}
	\label{EVM_fig}
\end{figure}
As the CU-ADMM algorithm can be roughly regarded as an optimizer that achieves the minimal EVM with only one execution,
 we can explore the relationship between EVM and PAPR.
From Fig. \ref{EVM_fig}, it can be observed that the EVM drops rapidly as the target PAPR increases.
When we set a higher PAPR,
the EVM deterioration is alleviated since fewer amplitudes of the signal need to be changed.
In addition, this method leads to a fair EVM performance for each subband.
With the help of CU-ADMM, we can wisely choose system parameters satisfying the PAPR and EVM requirements.

\section{Conclusions}
{\color{black}
In this paper, the PAPR reduction techniques are investigated for the mixed-numerology transmissions.
We first consider the classical clipping based approach and show that the direct application of the ICF method to composite signal will result in the INI accumulation.
By exploiting the clipping noise rather than the clipped signal, the NS-ICF method is able to reduce the high peaks without increasing the INI.
Next, we take the EVM distortion into consideration and 
formulate the task into an EVM minimization problem subject to the PAPR constraint.
Then, taking advantage of the separable structure of the convex programming, two low complexity algorithms, named as O-ADMM and CU-ADMM, are developed to efficiently solve this problem.  
As its decomposing and coordinating procedure can be associated with the structure of mixed numerology, the ADMM framework provides a tractable and scalable solution for the PAPR reduction problem with any number of numerologies.
Furthermore, the proposed PAPR reduction methods can be easily extended to the F-OFDM and W-OFDM. 
In combination with the filtering and windowing techniques, the proposed PAPR reduction methods achieve the suppression of OOBE as well as alleviate the spectrum regrowth caused by PA nonlinearity.

Our future work will focus on the joint optimization of the OOBE and PAPR using the ADMM framework.
For example, the combination of the optimization method and windowing technique discussed in Section V does not consider the windowing function as the optimization objective.
Hence, we can further take into account of other possible optimized designs. 
}

\bibliographystyle{IEEEtran}
\bibliography{IEEEabrv,Summary1221}

\begin{thebibliography}{10}
\providecommand{\url}[1]{#1}
\csname url@samestyle\endcsname
\providecommand{\newblock}{\relax}
\providecommand{\bibinfo}[2]{#2}
\providecommand{\BIBentrySTDinterwordspacing}{\spaceskip=0pt\relax}
\providecommand{\BIBentryALTinterwordstretchfactor}{4}
\providecommand{\BIBentryALTinterwordspacing}{\spaceskip=\fontdimen2\font plus
\BIBentryALTinterwordstretchfactor\fontdimen3\font minus
  \fontdimen4\font\relax}
\providecommand{\BIBforeignlanguage}[2]{{%
\expandafter\ifx\csname l@#1\endcsname\relax
\typeout{** WARNING: IEEEtran.bst: No hyphenation pattern has been}%
\typeout{** loaded for the language `#1'. Using the pattern for}%
\typeout{** the default language instead.}%
\else
\language=\csname l@#1\endcsname
\fi
#2}}
\providecommand{\BIBdecl}{\relax}
\BIBdecl

\bibitem{2017ICM-ZhangMulti}
L.~Zhang, A.~Ijaz, P.~Xiao, and R.~Tafazolli, ``Multi-service system: An
  enabler of flexible 5{G} air interface,'' \emph{IEEE Commun. Mag.}, vol.~55,
  no.~10, pp. 152--159, 2017.

\bibitem{2018ICSM-ZaidiOFDM}
\BIBentryALTinterwordspacing
A.~A. Zaidi, R.~Baldemair, V.~Moles-Cases, N.~He, K.~Werner, and A.~Cedergren,
  ``{OFDM} numerology design for 5{G} new radio to support {IoT}, {eMBB}, and
  {MBSFN},'' \emph{IEEE Communications Standards Magazine}, vol.~2, no.~2, pp.
  78--83, 2018.

\bibitem{2016IA-A.IjazEnabling}
A.~Ijaz, L.~Zhang, M.~Grau, A.~Mohamed, S.~Vural, A.~U. Quddus, M.~A. Imran,
  C.~H. Foh, and R.~Tafazolli, ``Enabling massive {IoT} in 5{G} and beyond
  systems: {PHY} radio frame design considerations,'' \emph{IEEE Access},
  vol.~4, pp. 3322--3339, 2016.

\bibitem{2016-TR38.913Study}
3GPP, ``Study on scenarios and requirements for next generation access
  technologies,'' tech. rep. 38.913, Tech. Rep. 4.0.0, Oct. 2016.

\bibitem{2016IJSAC-C.L.INew}
C.~I, S.~Han, Z.~Xu, S.~Wang, Q.~Sun, and Y.~Chen, ``New paradigm of 5{G}
  wireless internet,'' \emph{IEEE J. Sel. Areas Commun.}, vol.~34, no.~3, pp.
  474--482, 2016.

\bibitem{2017-3GPPTechnical}
3GPP, ``Technical specication group radio access network; {NR}; physical layer;
  general description ({R}elease 15),'' TS 38.201, Tech. Rep., 2017, v1.0.0.

\bibitem{2012Sahin}
A.~Sahin and H.~Arslan, ``Multi-user aware frame structure for {OFDMA} based
  system,'' in \emph{Proc. IEEE Veh. Technol. Conf. (VTC)}, Quebec, Canada,
  Sep. 2012, pp. 1--5.

\bibitem{2017IToWC-ZhangSubband}
L.~Zhang, A.~Ijaz, P.~Xiao, A.~Quddus, and R.~Tafazolli, ``Subband filtered
  multi-carrier systems for multi-service wireless communications,'' \emph{IEEE
  Trans. Wireless Commun.}, vol.~16, no.~3, pp. 1893--1907, 2017.

\bibitem{2017IJoSAiC-Guan5G}
P.~Guan, D.~Wu, T.~Tian, J.~Zhou, X.~Zhang, L.~Gu, A.~Benjebbour, M.~Iwabuchi,
  and Y.~Kishiyama, ``{5G} field trials: {OFDM}-based waveforms and mixed
  numerologies,'' \emph{IEEE J. Sel. Areas Commun.}, vol.~35, no.~6, pp.
  1234--1243, 2017.

\bibitem{2018JoMM-YazarFlexible}
A.~Yazar and H.~Arslan, ``Flexible multi-numerology systems for 5g new radio,''
  \emph{Journal of Mobile Multimedia}, vol.~14, no.~2, pp. 367--394, Oct. 2018.

\bibitem{2019EJoWCaN-YazarReliability}
\BIBentryALTinterwordspacing
------, ``Reliability enhancement in multi-numerology-based 5{G} new radio
  using {INI}-aware scheduling,'' \emph{EURASIP Journal on Wireless
  Communications and Networking}, vol. 110, no.~1, pp. 1--14, May 2019.

\bibitem{2009JCaN-D.-W.LimOverview}
D.~Lim, S.~Heo, and J.~No, ``An overview of peak-to-average power ratio
  reduction schemes for {OFDM} signals,'' \emph{J. Commun. Networks}, vol.~11,
  no.~3, pp. 229--239, 2009.

\bibitem{2017ICM-Lien5G}
S.-Y. Lien, S.-L. Shieh, Y.~Huang, B.~Su, Y.-L. Hsu, and H.-Y. Wei, ``5{G} new
  radio: Waveform, frame structure, multiple access, and initial access,''
  \emph{IEEE Commun. Mag.}, vol.~55, no.~6, pp. 64--71, 2017.

\bibitem{2013ICST-RahmatallahPeak}
\BIBentryALTinterwordspacing
Y.~Rahmatallah and S.~Mohan, ``Peak-to-average power ratio reduction in {OFDM}
  systems: A survey and taxonomy,'' \emph{IEEE Communications Surveys \&
  Tutorials}, vol.~15, no.~4, pp. 1567--1592, 2013.

\bibitem{1996EL-BaumlReducing}
R.~W. Bauml, R.~F.~H. Fischer, and J.~B. Huber, ``Reducing the peak-to-average
  power ratio of multicarrier modulation by selected mapping,'' \emph{Electron.
  Lett.}, vol.~32, no.~22, pp. 2056--2057, Oct 1996.

\bibitem{1997EL-MullerOFDM}
S.~H. Muller and J.~B. Huber, ``{OFDM} with reduced peak-to-average power ratio
  by optimum combination of partial transmit sequences,'' \emph{Electron.
  Lett.}, vol.~33, no.~5, pp. 368--369, 1997.

\bibitem{1998Tellado}
J.~Tellado and J.~M. Cioffi, ``Peak power reduction for multicarrier
  transmission,'' in \emph{IEEE GLOBECOM}, vol.~99, 1998, pp. 5--9.

\bibitem{2011ITC-M.SabbaghianShannon}
B.~S. M.~Sabbaghian, Y.~Kwak and V.~Tarokh, ``Near shannon limit and low peak
  to average power ratio turbo block coded {OFDM},'' \emph{IEEE Trans.
  Commun.}, vol.~59, no.~8, pp. 2042--2045, 2011.

\bibitem{2016ICM-ZhangWaveform}
X.~Zhang, L.~Chen, J.~Qiu, and J.~Abdoli, ``On the waveform for 5{G},''
  \emph{IEEE Commun. Mag.}, vol.~54, no.~11, pp. 74--80, 2016.

\bibitem{2002EL-ArmstrongPeak}
\BIBentryALTinterwordspacing
J.~Armstrong, ``Peak-to-average power reduction for {OFDM} by repeated clipping
  and frequency domain filtering,'' \emph{Electron. Lett.}, vol.~38, no.~5, pp.
  246--247, 2002.

\bibitem{2006ITSP-AggarwalMinimizing}
\BIBentryALTinterwordspacing
A.~Aggarwal and T.~H. Meng, ``Minimizing the peak-to-average power ratio of
  {OFDM} signals using convex optimization,'' \emph{IEEE Trans. Signal
  Process.}, vol.~54, no.~8, pp. 3099--3110, 2006.

\bibitem{2009IJSTSP-LiuError}
\BIBentryALTinterwordspacing
Q.~Liu, R.~J. Baxley, X.~Ma, and G.~T. Zhou, ``Error vector magnitude
  optimization for {OFDM} systems with a deterministic peak-to-average power
  ratio constraint,'' \emph{{IEEE} J. Sel. Topics Signal Process.}, vol.~3,
  no.~3, pp. 418--429, 2009.

\bibitem{2011ITC-WangOptimized}
\BIBentryALTinterwordspacing
Y.~Wang and Z.~Luo, ``Optimized iterative clipping and filtering for papr
  reduction of {OFDM} signals,'' \emph{IEEE Trans. Commun.}, vol.~59, no.~1,
  pp. 33--37, 2011.

\bibitem{2013ITC-ZhuSimplified}
\BIBentryALTinterwordspacing
X.~Zhu, W.~Pan, H.~Li, and Y.~Tang, ``Simplified approach to optimized
  iterative clipping and filtering for papr reduction of {OFDM} signals,''
  \emph{IEEE Trans. Commun.}, vol.~61, no.~5, pp. 1891--1901, 2013.

\bibitem{2010FTML-BoydDistributed}
S.~P. Boyd, N.~Parikh, E.~Chu, B.~Peleato, and J.~Eckstein, ``Distributed
  optimization and statistical learning via the alternating direction method of
  multipliers,'' \emph{Found. Trends Mach. Learn.}, vol.~3, p. 1–122, 2011.

\bibitem{2018ITVT-BaoADMM}
\BIBentryALTinterwordspacing
H.~Bao, J.~Fang, Q.~Wan, Z.~Chen, and T.~Jiang, ``An {ADMM} approach for {PAPR}
  reduction for large-scale {MIMO}-{OFDM} systems,'' \emph{IEEE Trans. Veh.
  Technol.}, pp. 1--1, 2018.

\bibitem{2018ITSP-YongchaoWangOptimized}
Y.~Wang, Y.~Wang, and Q.~Shi, ``Optimized signal distortion for {PAPR}
  reduction of {OFDM} signals with {IFFT/FFT} complexity via {ADMM}
  approaches,'' \emph{IEEE Trans. Signal Process.}, vol.~67, no.~2, pp.
  399--414, 2019.

\bibitem{-Non}
S.~Pagadarai, A.~Kliks, H.~Bogucka, and A.~M. Wyglinski, ``Non-contiguous
  multicarrier waveforms in practical opportunistic wireless systems,''
  \emph{IET Radar, Sonar Navigation}, vol.~5, no.~6, pp. 674--680, Jul. 2011.

\bibitem{2014IN-Z.KhanCarrier}
Z.~Khan, H.~Ahmadi, E.~Hossain, M.~Coupechoux, L.~A. Dasilva, and J.~J.
  Lehtomäki, ``Carrier aggregation/channel bonding in next generation cellular
  networks: methods and challenges,'' \emph{IEEE Network}, vol.~28, no.~6, pp.
  34--40, Nov. 2014.

\bibitem{2017-3GPPNR}
3GPP, ``{NR;} base station transmission and reception ({R}elease 15),'' TS
  38.104, Tech. Rep., 2017.

\bibitem{2001ITC-Ochiaidistribution}
\BIBentryALTinterwordspacing
H.~Ochiai and H.~Imai, ``On the distribution of the peak-to-average power ratio
  in {OFDM} signals,'' \emph{IEEE Trans. Commun.}, vol.~49, no.~2, pp.
  282--289, 2001.

\bibitem{2000IJSAC-OchiaiPerformance}
\BIBentryALTinterwordspacing
------, ``Performance of the deliberate clipping with adaptive symbol selection
  for strictly band-limited {OFDM} systems,'' \emph{IEEE J. Sel. Areas
  Commun.}, vol.~18, no.~11, pp. 2270--2277, 2000.

\bibitem{2008ITVT-WangAnalysis}
\BIBentryALTinterwordspacing
L.~Wang and C.~Tellambura, ``Analysis of clipping noise and tone-reservation
  algorithms for peak reduction in {OFDM} systems,'' \emph{IEEE Trans. Veh.
  Technol.}, vol.~57, no.~3, pp. 1675--1694, 2008.

\bibitem{CVX}
\BIBentryALTinterwordspacing
M.~Grant and S.~Boyd, \emph{CVX: Matlab software for disciplined convex
  programming (web page and software)}, Oct. 2008. [Online]. Available:
  \url{http://stanford.edu/ boyd/cvx}
\BIBentrySTDinterwordspacing

\bibitem{2004-BoydConvex}
S.~Boyd and L.~Vandenberghe, \emph{Convex Optimization}.\hskip 1em plus 0.5em
  minus 0.4em\relax Cambridge University Press, 2004.

\bibitem{lin2015global}
T.~Lin, S.~Ma, and S.~Zhang, ``On the global linear convergence of the admm
  with multiblock variables,'' \emph{SIAM J. Optim.}, vol.~25, no.~3, pp.
  1478--1497, 2015.

\bibitem{2016ICM-Waveform}
A.~A. Zaidi, R.~Baldemair, H.~Tullberg, H.~Bjorkegren, L.~Sundstrom, J.~Medbo,
  C.~Kilinc, and I.~D. Silva, ``Waveform and numerology to support 5{G}
  services and requirements,'' \emph{{IEEE} Communications Magazine}, vol.~54,
  no.~11, pp. 90--98, 2016.

\end{thebibliography}

\end{document}